\documentclass[a4paper,12pt]{article}
\usepackage{latexsym,amssymb}
\usepackage{amsmath,amssymb,amsxtra,amscd,amsthm}
\usepackage{amsfonts}
\usepackage[mathscr]{eucal}
\DeclareFontFamily{U}{rsfs}{\skewchar\font127 }
\DeclareFontShape{U}{rsfs}{m}{n}{
   <5> rsfs5
   <6> rsfs6
   <7> rsfs7
   <8> rsfs8
   <9> rsfs9
   <10> rsfs10
   <10.95> rsfs11
   <12> rsfs12
   <14.4> rsfs14
   <17.28> rsfs17
   <20.74> rsfs20
   <24.88> rsfs25
   <29.86-> rsfs30}{}
\DeclareMathAlphabet\scr{U}{rsfs}{m}{n}

\def\mZ{\mathbb{Z}}

\def\tL{\mathrm{L}}

\def\tS{\mathrm{S}}

\DeclareFontFamily{U}{rsf}{} \DeclareFontShape{U}{rsf}{m}{n}{
  <5> <6> rsfs5 <7> <8> <9> rsfs7 <10-> rsfs10}{}
\DeclareMathAlphabet\Scr{U}{rsf}{m}{n}

\def\bb1{\textup{\small{1}} \kern-3.8pt \textup{1}}

\def\SL2Z{\tS\tL(2,\mZ)}

\numberwithin{equation}{section}

\providecommand{\href}[2]{#2}

\newtheoremstyle{plain} 
  {5pt}
  {10pt}
  {\rmfamily} 
  {}
  {\scshape} 
  {}
  {\newline} 
  {}
\swapnumbers

\def\a{\alpha}
\def\b{\beta}

\newcommand{\SO}{\mathop{\rm SO}}
\newcommand{\U}{\mathop{\rm {}U}}

\newcommand{\Sp}{\mathop{\rm {}Sp}}
\newcommand{\OSp}{\mathop{\rm {}OSp}}

\newcommand{\so}{\mathfrak{so}}

\newcommand{\sym}{\mathfrak{sp}}

\newcommand{\osp}{\mathfrak{osp}}

\newcommand{\ft}[2]{{\textstyle\frac{#1}{#2}}}

\begin{document}
\begin{titlepage}
\title{
{\bf Theory of Superdualities and the Orthosymplectic Supergroup
}
~\\
\medskip
\author{Pietro Fr\'e$^1$,
Pietro Antonio Grassi$^2$, \\ Luca Sommovigo$^2$, and Mario Trigiante$^3$\\
~\\
{\small \it $^1$ Dipartimento di Fisica Teorica, Universit{\`a} di Torino, }\\
{\small\it
via P. Giuria 1, I-10125 Torino, Italy $\&$ INFN -
Sezione di Torino}\\
{\small\it $^2$ DISTA, Universit\`a del Piemonte Orientale, }\\
{\small\it Via T. Michel 11,  Alessandria, 15100, Italy
$\&$ INFN - Sezione di
Torino}
~\\
{\small\it $^3$ Dipartimento di Fisica, Politecnico di Torino,}\\
{\small\it C.so Duca degli Abruzzi, 24, I-10129 Torino, Italy $\&$ INFN -
Sezione di Torino}
}}
\maketitle
\vskip+10pt

\begin{abstract}
We study the dualities for sigma models with fermions and bosons.
We found that the generalization of the $\SO(m,m)$ duality for
$D=2$ sigma models and the ${\rm Sp}(2n)$ duality for $D=4$ sigma
models is the orthosymplectic duality $\OSp(m,m|2 n)$. We study
the implications of this and we derive the most general $D=2$
sigma model, coupled to fermionic and bosonic one-forms, with such
dualities. To achieve this we generalize Gaillard-Zumino analysis
to orthosymplectic dualities, which requires to define embedding
of the superisometry group of the target space into the duality
group. We finally discuss the recently proposed fermionic
dualities as a by-product of our construction.


\end{abstract}

\thispagestyle{empty}
\end{titlepage}
\tableofcontents
\newpage
\section{Introduction}
In the last twenty years dualities have played a major role in
understanding non-perturba\-tive aspects of superstring theory.
They have indeed unveiled relations between different
compactifications of superstring/M-theory, showing that such
realizations can be seen as different descriptions of a same
microscopic dynamics. It has also been conjectured long ago that
superstring dualities are encoded in the global symmetries of the
low-energy effective supergravity theory \cite{Hull:1994ys}.\par
Until recently dualities have been characterized as mappings
between bosonic backgrounds (bosonic dualities) which do not
affect the fermions. Among them, T-dualities
\cite{Buscher:1987qj,Giveon:1994fu} are mappings between
``large''-- and ``small''--radius compactifications of superstring
theory and are realized  as non-local redefinitions of the
world-sheet bosonic fields (i.e. coordinates on the target
space-time). A condition for such a redefinition to be feasible is
that the background moduli (supergravity fields) be independent
of the ``dualized'' coordinates. \par Berkovits and Maldacena, in
\cite{Berkovits:2008ic}, introduced a generalization of the
bosonic T-duality, called ``fermionic'' (or super-) T-duality,
which also involves the fermionic modes and which is realized as a
non-local redefinition of the world-sheet fermions (fermionic
coordinates in the background supermanifold). This duality can be
consistently defined on superstring backgrounds in which the
fields do not explicitly depend on the ``dualized''
super-coordinates.
 These ideas have been investigated so far mainly from the world-sheet point
 of view on specific space-time backgrounds \cite{Chandia:2009yv,Adam:2009kt,Beisert:2009cs}
 Therefore a general, background independent, characterization of ``fermionic'' dualities
  is still missing.\par In
the present paper we make progress in this direction by defining a
supersymmetric sigma model which is globally invariant with
respect to a \emph{super-duality} group.\par We consider a
globally supersymmetric sigma model in two dimensions, coupled to
a set of scalar and fermion one-forms
behaving as field strengths  of scalar and fermion ``0-form
fields''. The sigma model scalar fields span a supermanifold
$\mathcal{SM}^{(x|y)}$ of the form
$\widehat{\mathcal{G}}/\widehat{\mathcal{H}}$. We define,
generalizing the Gaillard-Zumino construction
\cite{Gaillard:1981rj}, the most general coupling of the
super-sigma model fields to the bosonic and fermionic one-forms
which allows to promote the super-isometry group
$\widehat{\mathcal{G}}$ of $\mathcal{SM}^{(x|y)}$ to global
super-symmmetry of the whole model, namely to a
\emph{super-duality}. This requires an embedding of
$\widehat{\mathcal{G}}$ in the supergroup $\OSp(m,m|4n)$, and of
$\widehat{\mathcal{H}}$ inside $\OSp(m|2n)\times \OSp(m|2n)$,
where $m$ and $2n$ are the numbers of bosonic and fermionic
one-forms respectively.
\par
The form of the sigma model is suggested by the Pure Spinor
Formulation of string theory \cite{Berkovits:2000fe} where the
bosonic and fermionic degrees of freedom are treated on the same
footing. The fields appearing in our sigma models are divided into
two sets of fields: the proper scalars and fermions parameterizing
a homogenous supermanifold and 0-form fields (part of which can be
interpreted as  Matzner-Missner fields). They are identified with
the coordinates of a superspacetime. On specific backgrounds, for
example on $AdS_p \times S^{10-p}$, part of the bosonic and
fermionic coordinates on the associated superspace, are treated as
the proper fields and the others, as the 0-form fields.

By coupling the model to supergravity background and eliminating
the auxiliary fields, one can recast the sigma model in the form discussed here. In that case, the
couplings acquire a physical interpretation.
\par The paper is organized as follows. In section
\ref{sec1} we consider the case of a bosonic sigma model in
$D=2\,p$ dimensions, with a homogeneous target space of the form
$\mathcal{G}/\mathcal{H}$, coupled to a number $n$ of $p$-form
field strengths, and review the Gaillard-Zumino construction of
the general form of this coupling which allows to promote the
isometry group $\mathcal{G}$ to a global on-shell symmetry of the
full model. This construction requires the isometry group
$\mathcal{G}$ to be embedded in the groups $\Sp(2n,\mathbb{R})$
and $\SO(n,n)$ for even and odd $p$ respectively. We will than
specialize  our discussion, in Section \ref{sec2}, to $D=2$ and
consider a generic bosonic sigma model coupled to one-form field
strengths, which may be seen as originating from a dimensional
reduction of the bosonic sector of a $D=4$ supergravity, as it is
explained in Section \ref{sec3}. In addition, we discuss the
relation between our sigma model and the Green-Schwarz sigma model
on a given supergravity background. In the last section, we
further elaborate on this identification.

In Section \ref{sec4}, we extend this analysis to a
$D=2$ super-sigma model coupled to a generic number of scalar and
fermion one-form field strengths ($p=1$). Our construction in this
more general setting requires the sigma model super-manifold
$\widehat{\mathcal{G}}/\widehat{\mathcal{H}}$ to be embedded in
$\OSp(m,m|4n)/[\OSp(m|\,2n)\times \OSp(m|\,2n)]$.
 In Section \ref{sec5} we explicitly define such embedding on the case in
which
$\mathcal{SM}^{(x|y)}=\frac{OSp(p,p|4r)}{\SO(p)\times\SO(p)\times
\U(2 r)}\times \frac{OSp(q,q|2s)}{\SO(q)\times\SO(q)\times \U(s)}$
and the number of bosonic and fermionic one-forms are $m=2 pq+4 rs$
and $n=2 ps+4qr$ repsectively. In Section \ref{sec6} we discuss in
detail the action of the superdualities on the fields of the model
and work out an explicit example.

\section{Electric/magnetic dualities in  $D=4$ and $D=2$ bosonic field
theories}\label{sec1} We first review the bosonic set up of
electric/magnetic duality rotations in D=4 and D=2. In both cases,
the relevant Lagrangian is constructed with a set of scalars
parameterizing a coset manifold $\mathcal{G}/\mathcal{H}$ to which
we add with a set of $\frac{D-2}{2}$-forms  (vector fields for
D=4, ``0-form'' scalars for D=2) on whose electric/magnetic field
strengths (two-forms for D=4, one-forms for D=2) the group
$\mathcal{G}$ acts by means of linear transformations (symplectic
in the D=4 case, pseudo-orthogonal in D=2 case). Hence, given an
element $g$ of the isometry group $\mathcal{G}$, there must be an
embedding $g \rightarrow \Lambda_g \in \mathrm{Sp}(2n_V,
\mathbb{R})$ for $D=4$ and $g \rightarrow \Lambda_g \in
\mathrm{SO}(n_S, n_S)$ for D=2, which leads to the construction of
the kinetic terms for the vectors in D=4 and for the 0-form
scalars in D=2 by the using the so called Gaillard-Zumino (GZ)
formula. The transformations of $\mathrm{Sp}(2n_V, \mathbb{R})$
and of $\mathrm{SO}(n_S, n_S)$, which are not in the image  of
$\mathcal{G}$ through this embedding, correspond to the possible
non-trivial duality transformations leading to new theories.
\par
\subsection{$D=4$ bosonic supergravity and its dualities}
\label{d4sugrasym} In  $D=4$ \textit{ungauged supergravity}, with
$N_Q$ supercharges, the bosonic Lagrangian admits the following
general form
\begin{eqnarray}
\mathcal{L}^{(4)} &=& \sqrt{\mbox{det}\, g}\left[-2R[g] -
\frac{1}{2}
\partial_{{\mu}}\phi^a\partial^{{\mu}}\phi^b h_{ab}(\phi) \,
+ \,
\mbox{Im}\mathcal{N}_{\Lambda\Sigma}\, F_{{\mu}{\nu}}^\Lambda
F^{\Sigma|{\mu}{\nu}}\right] \nonumber\\
&+&
\frac{1}{2}\mbox{Re}\mathcal{N}_{\Lambda\Sigma}\, F_{{\mu}{\nu}}^\Lambda
F^{\Sigma}_{{\rho}{\sigma}}\epsilon^{{\mu}{\nu}{\rho}{\sigma}}
\label{d4generlag}
\end{eqnarray}
where $\phi^a$ denotes the whole set of $n_S$ scalar fields
parameterizing the scalar manifold $ \mathcal{M}_{scalar}^{D=4}$
which, for $N_Q > 8$, is necessarily a coset manifold:
\begin{equation}
  \mathcal{M}_{scalar}^{D=4} \, =
  \,\frac{\mathcal{G}}{\mathcal{H}}
\label{cosettoquando}
\end{equation}
For $N_Q \le 8$, condition (\ref{cosettoquando}) is not implied by
supersymmetry. However  $\mathcal{N}=2$ supergravity, i.e. for
$N_Q =8$, a large variety of homogeneous special K\"ahler
manifolds fall into the set up of the present general discussion.
The fields $\phi^a$ have $\sigma$--model interactions dictated by
the metric $h_{ab}(\phi)$ of $\mathcal{M}_{scalar}^{D=4}$.
\par
The theory includes also $n_V=n$ vector
fields $A_{{\mu}}^\Lambda$ for which
\begin{eqnarray}
  \mathcal{F}^{\pm| \Lambda}_{{\mu}{\nu}} \equiv \ft 12
  \left({F}^{\Lambda}_{{\mu}{\nu}} \pm \, {\rm i} \,\star F_{\mu\nu}
  \right)\,\,,\mbox{where}\,\,\,\,\,\,\, \star F_{\mu\nu}\equiv\frac{\sqrt{\mbox{det}\, g}}{2}
  \epsilon_{{\mu}{\nu}{\rho}{\sigma}} \,
  F^{{\rho}{\sigma}}\,,
\label{Fpiumeno}
\end{eqnarray}
denote the self-dual (respectively antiself-dual) parts of the
field-strengths: ${}^\star \mathcal{F}^{\pm| \Lambda}=\mp {\rm i}
\,\mathcal{F}^{\pm| \Lambda}$. As displayed in
eq.(\ref{d4generlag}) they are non-minimally coupled to the
scalars via the symmetric complex matrix\footnote{Note that, in
this conventions, the physical domain of the kinetic terms for the
vector fields is defined by the condition ${\rm Im}\mathcal{N}<0
$.}
\begin{equation}
  \mathcal{N}_{\Lambda\Sigma}(\phi)
  ={\rm i}\, \mbox{Im}\mathcal{N}_{\Lambda\Sigma}+ \mbox{Re}\mathcal{N}_{\Lambda\Sigma}
\label{scriptaenna}
\end{equation}
  Following the notations and
the conventions of \cite{myparis}, it can be shown that the
isometry group $\mathcal{G}$, global symmetry of the sigma model
action, can be promoted to global on-shell symmetry of the theory,
provided the following symplectic embedding of $\mathcal{G}$ is
defined:
\begin{equation}
 \boxed{ \mathcal{G} \mapsto \mathrm{Sp(2n, \mathbb{R})} }\quad ; \quad
  n = n_V \, \equiv \, \mbox{$\#$ of vector fields}
\label{sympembed}
\end{equation}
which associates with each element of $\mathcal{G}$ a symplectic
electric-magnetic duality transformation on the field strengths
${F}^{ \Lambda}_{\mu\nu}$ plus their magnetic duals. The latter
therefore define a symplectic representation $\mathbf{W}$ of
$\mathcal{G}$.\par
 More specifically, the embedding (\ref{sympembed}) implies that each element $\xi
\in \mathcal{G}$ is represented by means of a suitable real
symplectic matrix:
\begin{equation}
  \xi \mapsto \Lambda_\xi \equiv \left( \begin{array}{cc}
     A_\xi & B_\xi \\
     C_\xi & D_\xi \
  \end{array} \right)
\label{embeddusmatra}
\end{equation}
satisfying the defining relation:
\begin{equation}
  \Lambda_\xi ^T \, \left( \begin{array}{cc}
     \mathbf{0}_{n \times n}  & { \mathbf{1}}_{n \times n} \\
     -{ \mathbf{1}}_{n \times n}  & \mathbf{0}_{n \times n}  \
  \end{array} \right) \, \Lambda_\xi = \left( \begin{array}{cc}
     \mathbf{0}_{n \times n}  & { \mathbf{1}}_{n \times n} \\
     -{ \mathbf{1}}_{n \times n}  & \mathbf{0}_{n \times n}  \
  \end{array} \right)
\label{definingsympe}
\end{equation}
which implies the following relations on the $n \times n$ blocks:
\begin{eqnarray}
A^T_\xi  \, C_\xi  - C^T_\xi  \, A_\xi  & = & 0 \nonumber\\
A^T_\xi  \, D_\xi  - C^T_\xi  \, B_\xi  & = & \mathbf{1}\nonumber\\
B^T_\xi  \, C_\xi  - D^T_\xi  \, A_\xi & = & - \mathbf{1}\nonumber\\
B^T_\xi  \, D_\xi  - D^T_\xi  \, B_\xi  & =  & 0
\label{symplerele}
\end{eqnarray}
Under an element of the duality group the field strengths transform
as follows:
\begin{equation}
  \left(\begin{array}{c}
     \mathcal{F}^\pm \\
     \mathcal{G}^\pm \
  \end{array} \right)  ^\prime \, = \,\left( \begin{array}{cc}
     A_\xi & B_\xi \\
     C_\xi & D_\xi \
  \end{array} \right) \,  \left(\begin{array}{c}
     \mathcal{F}^\pm \\
     \mathcal{G}^\pm \
  \end{array} \right) \,,
\label{lucoidale1}
\end{equation}
where, by their own definitions:
\begin{equation}
    \mathcal{G}^+ = \mathcal{N} \, \mathcal{F}^+ \quad ; \quad \mathcal{G}^- = \overline{\mathcal{N}} \,
    \mathcal{F}^-
\label{lucoidale2}
\end{equation}
Consistency of eq. (\ref{lucoidale2}) with the transformation
(\ref{lucoidale1}) is guaranteed by the symplectic property of
$\Lambda_\xi$, provided the complex symmetric matrix $\mathcal{N}$
transforms as follows:
\begin{equation}
  \mathcal{N}^\prime = \left(  C_\xi  + D_\xi  \, \mathcal{N}\right) \, \left( A_\xi  + B_\xi  \,\mathcal{N}\right)
  ^{-1}
\label{Ntransfa}
\end{equation}
The condition (\ref{sympembed}), which defines the duality action
of the isometry group of the scalar manifold, also holds in $D>4$
even dimensions, with $D/2$ even as well. In this case $n$ denotes
the number of $(D-2)/2$-forms (vector fields in $D=4$ and
rank-three antisymmetric tensor fields in $D=8$). For $D$-even but
$D/2$ odd,  $n$ still refers to the number of $(D-2)/2$-forms
(scalar fields in $D=2$, rank-two antisymmetric tensor fields in
$D=6$ and rank-four antisymmetric tensor fields in $D=10$), but
the action on their field strengths and their duals is defined by
the embedding of $\mathcal{G}$ in the pseudo-orthogonal group
$\SO(n,n)$.
\subsubsection{Symplectic ${\rm Sp}(2m, \mathbb R)$ embeddings and
the Gaillard-Zumino formula for the period matrix $\mathcal{N}$}
Focusing on the isometry group of the canonical metric defined on
$\frac{\mathcal{G}}{\mathcal{H}}$, we must consider the embedding:
\begin{equation}
\boxed{\iota_\delta : \,  \mathcal{G} \, \longrightarrow \,
\mathrm{Sp}(2 n, \mathbb{R})} \label{embediso}
\end{equation}
This is an homomorphism between  finite dimensional Lie groups and
as such it can be determined explicitly. What we just need to know
is the dimension of the symplectic group, namely the number $ n$
of $\frac{D-2}{2}$--forms appearing in the theory. In $D=4$, for
example $n$ coincides with the number of vector fields $n_V$,
without supersymmetry, the dimension $n_S$ of the scalar manifold
\footnote{We restrict here and in the following to the manifold
spanned by the scalar fields which couple to the vectors through
the kinetic  matrix $\mathcal{N}$.} (namely the possible choices
of $\frac{{\cal G}}{{\cal H}}$) and the number of vectors $n_V$
are unrelated so that the possibilities covered by
eq.~(\ref{embediso}) are infinitely many. In supersymmetric
theories with number of the supercharges is greater or equal to
eight, instead, the two numbers $n_S$ and $n_V$ are related ($n_S$
being the number of scalar fields in the same supermultiplets as
the vector fields), so that there are finitely many cases to be
studied corresponding to the possible embeddings of given groups
$\mathcal{G}$ into a symplectic group $\mathrm{Sp}(2n,
\mathbb{R})$ of a given $n$.\par
 Apart from the details of the
specific case considered once a symplectic embedding is given
there is a general formula one can write down for the {\it period
matrix} {}\footnote{The matrix $\mathcal{N}$ is named period matrix because of its meaning when
the considered four-dimensional supergravity is obtained by compactification of 10D type IIB supergravity
on a Calabi-Yau three-fold. In that case the matrix $\mathcal{N}$, in full analogy with the period matrix of Riemann surfaces, is obtained by considering the matrix whose entries are the periods of the cohomology three-forms on a basis of homology three-cycles. The symplectic transformations are those
which leave invariant the intersection matrix of the three-cycles.}  ${\cal N}$
 that guarantees symmetry (${\cal N}^T = {\cal
N}$) and the required transformation properties.
\par
The real symplectic group $\mathrm{Sp}(2n ,\mathbb{R})$ is defined as the
set of all {\it real} $2n \times 2n$ matrices with an algebraic condition
\begin{equation}
\Lambda \, = \, \left (
\begin{array}{cc}
A & B \\
C & D
\end{array}
\right )\,, \hspace{2cm}
\Lambda^T \, \mathbb{C} \, \Lambda \, = \, \mathbb{C} \label{condiziona}
 \end{equation}
 where
\begin{equation}
\mathbb{C}  \, \equiv \, \left (
\begin{array}{cc} {\bf 0} & {\bf 1} \cr -{\bf 1} &
{\bf 0} \end{array} \right ) \label{definizia}
\end{equation}
We can change the basis of the fundamental symplectic
representation to a complex one, defined by the action of the
Cayley matrix:
\begin{equation}
{\cal C}_{\mathrm{Sp}} \, \equiv \, {\frac{1}{\sqrt{2}}} \, \left (
\begin{array}{cc} {\bf 1} & {\rm i}{\bf 1} \cr {\bf 1} & -{\rm
i}{\bf 1}
 \end{array} \right )
\label{cayley}
\end{equation}
which is adapted for $\mathrm {Sp}$ groups.
In this new basis  the generic  symplectic matrix $\Lambda$ in
(\ref{condiziona}) will become a complex matrix
$\mathcal{S}$ defined as follows:
\begin{eqnarray}
{\cal S}\, &\equiv& \mathcal{C}\,\Lambda\,\mathcal{C}^{-1}\,= \,
\left (
\begin{array}{cc} T & V^\star \cr V & T^\star \end{array} \right
)\,,\label{blocusplet}
\end{eqnarray}
where
\begin{eqnarray}
 T &=& {\frac{1}{2}}\, \left ( A - {\rm i} B
\right ) + {\frac{1}{2}}\, \left ( D + {\rm i} C \right ) \quad ;
\quad V = {\frac{1}{2}}\, \left ( A - {\rm i} B \right ) -
{\frac{1}{2}}\, \left ( D + {\rm i} C \right )\,. \label{mappetta}
\end{eqnarray}
In this complex basis the $\U(n)$ transformations inside
$\Sp(2n,\mathbb{R})$ have a block diagonal form:
\begin{eqnarray}
h&\in&\U(n)\,\,\,\,;\,\,\,\,\,\,h=\left(\begin{matrix}W & {\bf
0}\cr {\bf 0} & W^\star
\end{matrix}\right)\,\,\,,\,\,\,\,
W\,W^\dagger = W^\dagger \,W={\bf 1}\,.
\end{eqnarray}
The basic idea, to obtain the general formula for the matrix ${\cal
N}$, is that the symplectic embedding of the isometry group
$\mathcal{G}$ will be such that the isotropy subgroup
${\mathcal{H}}\subset \mathcal{G}$ gets embedded into the
maximal compact subgroup ${\mathrm{U}}(n)$, namely:
\begin{eqnarray}
\mathcal{G} & {\stackrel{\iota_\delta}{\longrightarrow}} &
\Sp(2n,\mathbb{R})\,; \quad {\mathcal{H}}
{\stackrel{\iota_\delta}{\longrightarrow}}  \mathrm{U}(n) \subset
\Sp(2n,\mathbb{R}) \label{gruppino}
\end{eqnarray}
If this condition is realized let $\mathbb{L}(\phi)$ be a
parametrization of the coset $\frac{\mathcal{G}}{\mathcal{H}}$
by means of coset representatives. By this we mean the following.
Let $\phi^I$ be local coordinates on the manifold
$\frac{\mathcal{G}}{{\mathcal{H}}}$: To each point $\phi \in
\frac{\mathcal{G}}{\mathcal{H}}$ we assign an element
$\mathbb{L}(\phi) \in \mathcal{G}$ in such a way that if
$\phi^\prime \ne \phi$, then no $h \in \mathcal{H}$ can exist such
that $\mathbb{L}(\phi^\prime)=\mathbb{L}(\phi)\cdot h$. In other
words for each equivalence class of the coset (labelled by the
coordinate $\phi$) we choose one representative element
$\mathbb{L}(\phi)$ of the class.  Relying on the symplectic
embedding of eq.(\ref{gruppino}) we obtain the symplectic
representation of the coset representative in the real basis:
\begin{eqnarray}
\mathbb{L}(\phi)\, \longrightarrow\, \mathcal{S}_{\Sp}(\phi)&=&\left(\begin{matrix}{ A}(\phi) &{
B}(\phi)\cr {C}(\phi) &{D}(\phi)
\end{matrix}\right)\,,
\end{eqnarray}
Since the coset representative is acted from the right and from
the left by elements of different groups, namely
$\mathcal{G}$ and $\mathcal{H}$ respectively, we may use
different bases for the the two indices of its matrix
representation and choose the real basis for the rows and the
complex one for the columns. We then define the following
mixed representation of $\mathbb{L}(\phi)$:
\begin{eqnarray}
\mathcal{S}_{\mathrm{mr}}(\phi)&\equiv &\left(\begin{matrix}\mathbf{ f}(\phi)
&\mathbf{ f}(\phi)^\star\cr \mathbf{ h}(\phi) &\mathbf{
h}(\phi)^\star
\end{matrix}\right)=\mathcal{S}_{\Sp}(\phi)\,\mathcal{C}^{-1}_{\Sp}\,,\label{mixarepra1}
\end{eqnarray}
where
\begin{eqnarray}
\mathbf{ f}(\phi)&=&\frac{1}{\sqrt{2}}\,\left({ A}(\phi)-{\rm
i}\,{ B}(\phi) \right)\,\,\,\,,\,\,\,\,\,\mathbf{
h}(\phi)=\frac{1}{\sqrt{2}}\,\left({C}(\phi)-{\rm i}\,{ D}(\phi)
\right)\,. \label{darstel}
\end{eqnarray}
From the  relations between the real blocks $A,B,C,D$ of a
symplectic matrix, expressed by the general equations
(\ref{symplerele}), one can verify that the $\mathbf{ f}$ and
$\mathbf{ h}$ blocks satisfy the following conditions:
\begin{eqnarray}\label{fhrels}
-{\rm
i}\,{\bf 1}&=&{\bf f}^\dagger\,{\bf h}-{{\bf h}}^\dagger\,{\bf
f}\,\,\,,\,\,\,\,{\bf f}^T\,{\bf h}-{{\bf h}}^T\,{\bf f}={\bf
0}\,, \\
({\bf f}\,{\bf f}^\dagger)^T&=&{\bf f}\,{\bf f}^\dagger\,\,\,,\,\,\,\,({\bf h}\,{\bf h}^\dagger)^T=
{\bf h}\,{\bf h}^\dagger\,\,\,,\,\,\,\,\, {\bf f}\,{\bf h}^\dagger-\bar{{\bf f}}\,{\bf h}^T={\rm i}\,{\bf 1}\,.
\nonumber
\end{eqnarray}
The action of an isometry $\xi\in \mathcal{G}$, represented by the real
symplectic matrix $\Lambda_\xi $ defined in eq.(\ref{embeddusmatra}),
on a point $\phi^I$ of the manifold is then described as follows:
\begin{eqnarray}
\Lambda_\xi\,\mathcal{S}_{\mathrm{mr}}(\phi)&=&\mathcal{S}_{\mathrm{mr}}(\xi(\phi))\,\left(\begin{matrix}W(\xi,\phi)
& {\bf 0}\cr {\bf 0} & W(\xi,\phi)^\star
\end{matrix}\right)\,, \label{cosettone}
\end{eqnarray}
where $\xi(\phi)$ denotes the image of the point $\phi$ through
$\xi$ and $W(\xi,\phi)$ is a suitable $\U(n)$ compensator
depending on both $\xi$ and $\phi$. Combining eq.s
(\ref{cosettone}),(\ref{darstel}), with eq (\ref{embeddusmatra})
we immediately obtain:
\begin{eqnarray}
\mathbf{f}(\xi(\phi))&=&\left[A_\xi\,\mathbf{f}(\phi)+B_\xi\,\mathbf{h}(\phi)\right]\,W(\xi,\phi)^\star\,
,\nonumber\\
\mathbf{h}(\xi(\phi))&=&\left[C_\xi\,\mathbf{f}(\phi)+D_\xi\,\mathbf{h}(\phi)\right]\,W(\xi,\phi)^\star\,,
\end{eqnarray}
If we define the $n\times n$ matrix $\mathcal{N}$ as follows:
\begin{eqnarray}
\mathcal{N}(\phi)&\equiv
&\mathbf{h}(\phi)\,\mathbf{f}(\phi)^{-1}=
\left [C(\phi) - i D(\phi)\right ] \left [ A(\phi) - i B(\phi)\right]^{-1}\,,\label{masterformula}
\end{eqnarray}
it is straightforward to verify that, under a generic isometry
$\xi$, it transforms as in eq. (\ref{Ntransfa}).
\par
It is also an immediate consequence of the last of eq.s
(\ref{fhrels}), satisfied by $\mathbf{f}$ and $\mathbf{h}$, that
the matrix in eq.(\ref{masterformula}) is symmetric
\begin{equation}
{\cal N}^T \, = \, {\cal N} \label{massi}
\end{equation}
Eq. (\ref{masterformula}) is the master-formula derived  by
Gaillard and Zumino \cite{Gaillard:1981rj}. It explains the structure of
the gauge field kinetic terms in all $\mathcal{N}\ge 3$ extended
supergravity theories and also in those $\mathcal{N}=2$ theories where the
{\it special K\"ahler manifold} ${\cal SM}$ is a homogeneous of the form
${\cal G}/{\cal H}$.  In the following, we will
derive the same formula for the orthogonal embeddings.\par
Notice that, at the origin of the manifold, $\phi=0$,
$\mathcal{S}_{\rm Sp}$ is the $2n\times 2n$ identity matrix, namely
$A(0)=D(0)={\bf 1}$ and $B(0)=C(0)={\bf 0}$. From
(\ref{masterformula}) and (\ref{darstel}) we find that
$\mathcal{N}(0)=-{\rm i}\,{\bf 1}$. One can verify that ${\rm
Im}(\mathcal{N}(\phi))$ is negative definite for any $\phi$.
\subsection{$D=2$ bosonic sigma model and its dualities}
Let us now consider the general form of a Lagrangian in $D=2$.
Here we have two types of scalars, namely the   proper scalars
$\phi^a$ and the \textit{twisted scalars} or \textit{scalar
0-forms} $\pi^\alpha$. This distinction is important. The proper
scalars appear in the Lagrangian under the form of a usual
$\sigma$-model, as coordinates on the target manifold ${\cal
G}/{\cal H}$, while the {scalar 0-forms} appear only covered by
derivatives and in two terms, one symmetric, one antisymmetric.
The coefficients of these two terms are matrices depending on the
proper scalars. Explicitly the Lagrangian has the form (see
\cite{Fre:1995dw} for a general review) in the conformal gauge
$g_{\mu\nu} = \delta_{\mu\nu}$:
\begin{eqnarray}
S_{(D=2)} & = & \int \, d^2x  \left \{
- \ft 1 2 \, h_{ab}(\phi) \partial_\mu \phi^a \partial _\mu \phi^b
\right. \nonumber\\
\null & \null & \left.
+   \, \ft 12 \,\kappa \, \left [-
\partial_\mu \pi^\alpha \,  \gamma_{\alpha\beta}(\phi) \, \partial_\mu \pi^\alpha \,
+ \partial_\mu \pi^\alpha \, \theta_{\alpha\beta}(\phi)  \, \partial_\nu \pi^\beta \, \epsilon^{\mu\nu} \right]\right\}
\label{d2generlag}
\end{eqnarray}
where $\kappa$ is a normalization parameter that can always be
reabsorbed into the definition of the $0$-forms $\pi^\alpha$
and where, according to the general theory for the dimensions $D=4\nu + 2$
(see section 2.4 of \cite{myparis})
if $\mathcal{G}$ is the isometry group of the
$\sigma$-model metric $h_{ab}(\phi)$, then there is a
pseudo-orthogonal embedding:
$$\boxed{
  \mathcal{G} \mapsto \mathrm{SO(m,m) }}
$$
where $m$ is the total number of the scalar-forms $\pi^\alpha$.
 Hence for each element $\xi \in \mathcal{G}$ we have its
representation by means of a  suitable pseudo-orthogonal matrix:
\begin{equation}
  \xi \mapsto \Lambda_\xi \equiv \left( \begin{array}{cc}
     \mathcal{A}_\xi & \mathcal{B}_\xi \\
     \mathcal{C}_\xi & \mathcal{D}_\xi \
  \end{array} \right)
\label{embeddusmatra2o1}
\end{equation}
which satisfies the defining equation:
\begin{equation}
  \Lambda_\xi ^T \, \left( \begin{array}{cc}
     \mathbf{0}_{m \times m}  & { \mathbf{1}}_{m \times m} \\
     { \mathbf{1}}_{m \times m}  & \mathbf{0}_{m \times m}  \
  \end{array} \right) \, \Lambda_\xi = \left( \begin{array}{cc}
     \mathbf{0}_{m \times m}  & { \mathbf{1}}_{m \times m} \\
     { \mathbf{1}}_{m \times m}  & \mathbf{0}_{m \times m}  \
  \end{array} \right)
\label{definingorto}
\end{equation}
 implying the following relations on the $m \times m$ blocks:
\begin{eqnarray}
\mathcal{A}^T_\xi \, \mathcal{C}_\xi + \mathcal{C}^T_\xi \, \mathcal{A}_\xi & = & 0 \nonumber\\
\mathcal{A}^T_\xi \, \mathcal{D}_\xi + \mathcal{C}^T_\xi \, \mathcal{B}_\xi & = & \mathbf{1}\nonumber\\
\mathcal{B}^T_\xi \, \mathcal{C}_\xi + \mathcal{D}^T_\xi \, \mathcal{A}_\xi& = &  \mathbf{1}\nonumber\\
\mathcal{B}^T_\xi \, \mathcal{D}_\xi + \mathcal{D}^T_\xi \, \mathcal{B}_\xi & =  & 0
\label{pseudoerele}
\end{eqnarray}
Let us now introduce the $D=2$ analogue of the $D=4$ period matrix $\mathcal{N}$. It is the
following $ m \times m$ matrix:
\begin{equation}
  \mathcal{M} \equiv \theta + \gamma
\label{coulor}
\end{equation}
which also deserves the name of period matrix. Indeed if we were to think of the considered $D=2$ theory
as the result of a compactification of $D=10$ supergravity on a Calabi-Yau four-fold, $\mathcal{M}$ could be interpreted as the matrix of periods of the cohomology four-forms on a basis of homology four-cycles.
The pseudo-orthogonal character arises  from preservation of the intersection matrix of such cycles which in this case is symmetric rather than antisymmetric.
Under the group $\mathcal{G}$ the matrix $\mathcal{M}$ transforms as follows:
\begin{equation}
  \mathcal{M}^\prime = \left( \mathcal{ C}_\xi + \mathcal{D }_\xi\, \mathcal{M}\right) \,
    \left( \mathcal{A}_\xi + \mathcal{B}_\xi \,\mathcal{M}\right)
  ^{-1}
\label{Mtransfa2}
\end{equation}
\begin{equation}
 - {\mathcal{M}^T}^\prime = \left( \mathcal{ C}_\xi - \mathcal{D }_\xi\, \mathcal{M}^T\right) \,
    \left( \mathcal{A}_\xi -\mathcal{B}_\xi \,\mathcal{M}^T \right)
  ^{-1}
\label{Mtransfa2bis}
\end{equation}
\subsubsection{Pseudo-orthogonal $\mathrm{SO(m,m)}$ embeddings and the Gaillard-Zumino formula for the period matrix $\mathcal{M}$}
Let us now repeat the Gaillard-Zumino construction of the kinetic matrix $\mathcal{M}$ for the case of pseudo-orthogonal embeddings. The maximal non-compact real section of the $D_m$ Lie algebra is $\so(m,m)$ which exponentiates to the group $\mathrm{SO(m,m)}$.
\par
In the symplectic case, relevant to $D=4$ theories,
we used two bases, related by a Cayley transformation, where the matrices  of both the group and  the algebra were either \textit{symplectic real} or \textit{symplectic complex and pseudo-unitary} at the same time.  The reason for this double presentation of the group/algebra elements was that the two bases have complementary virtues. In the former,  the  duality rotations are simply and linearly realized on the electric-magnetic field strengths, yet the maximal compact subalgebra of the duality-algebra is not realized by block-diagonal matrices. In the latter, the maximal compact subalgebra is block-diagonal, but the action on the physical field strengths is not the simplest.
\par
The same situation occurs in the pseudo-orthogonal case, relevant
for $D=2$ theories. Also here it is convenient to use two bases,
the first optimizing the form of the electric-magnetic duality
rotations, the second in which the maximal compact subalgebra
$\so(\mathrm{m}) \oplus \so(\mathrm{m}) \subset \so(\mathrm{m,m})$
has a  block-diagonal representation. The difference is that in
the $\so(\mathrm{m,m})$ case both bases are real and the Cayley
transformation relating  them is also a real matrix. This goes
hand in hand with the fact that the period matrix $\mathcal{M}$ of
$D=2$ twisted scalars is a real matrix while the kinetic matrix of
$D=4$ one-forms  is a symmetric complex matrix. In the $D=4$ case
the separation between generalized coupling constants and
generalized theta-angles is performed by singling out the real and
imaginary part of $\mathcal{N}$. In $D=2$ the same separation is
performed by splitting $\mathcal{M}$ into its  symmetric and
antisymmetric parts.
\par
The two bases are defined by giving the form of the
pseudo-orthogonal invariant metric $\mathbb{C}$. In the first
basis, which we name \textit{off-diagonal}, $\mathbb{C}$  has the
following appearance:
\begin{equation}\label{offamatra}
    \mathbb{C}_{\mathrm{off}} \, = \, \left (
\begin{array}{cc} {\bf 0} & {\bf 1} \cr {\bf 1} &
{\bf 0} \end{array} \right ) \, \equiv \, \left( \begin{array}{cc}
     \mathbf{0}_{m \times m}  & { \mathbf{1}}_{m \times m} \\
     { \mathbf{1}}_{m \times m}  & \mathbf{0}_{m \times m}  \
  \end{array} \right)
\end{equation}
In the second basis, named \textit{diagonal}, the invariant matrix is the following:
\begin{equation}\label{diamatra}
  \mathbb{C}_{\mathrm{dia}}  \left ( \begin{array}{cc}
{\bf 1} & {\bf 0} \cr {\bf 0} & -{\bf 1}
 \cr \end{array} \right )\, \equiv \, \left( \begin{array}{cc}
     \mathbf{1}_{m \times m}  & { \mathbf{0}}_{m \times m} \\
     { \mathbf{0}}_{m \times m}  & - \, \mathbf{1}_{m \times m}  \
  \end{array} \right)
\end{equation}
The change  from the off-diagonal to the diagonal basis is performed by the following generalized Cayley matrix:
\begin{equation}
{\cal C}_{\mathrm{so}}\, \equiv \, {\frac{1}{\sqrt{2}}} \, \left ( \begin{array}{cc} {\bf 1}
& {\bf 1} \cr {\bf 1} & -{\bf 1}
 \end{array} \right ) \quad ; \quad {\cal C}_{\mathrm{so}}\, {\cal C}_{\mathrm{so}} \, = \, \mathbf{1}
 \quad ; \quad {\cal C}_{\mathrm{so}}\, = \, {\cal C}_{\mathrm{so}}^T \,
\label{cayley2}
\end{equation}
which satisfies the relation:
\begin{equation}\label{cambius}
    {\cal C}_{\mathrm{so}}\, \mathbb{C}_{\mathrm{off}}  \, {\cal C}_{\mathrm{so}}\, = \, \mathbb{C}_{\mathrm{dia}} \end{equation}
Let us define elements of the $\mathrm{SO(m,m)}$ group and of the $\so(\mathrm{m,m})$ Lie algebra in the off-diagonal basis:
\begin{equation}\label{sommdefi}
    \begin{array}{ccccccc}
       \mathrm{SO(m,m)} \, \supset \, \Lambda & = & \left (\begin{array}{cc}
\mathcal{A} & \mathcal{B} \\
\mathcal{C} & \mathcal{D}
\end{array}\right ) & \Rightarrow & \Lambda^T \, \mathbb{C}_{\mathrm{off}} \, \Lambda & = & \mathbb{C}_{\mathrm{off}} \\
        \null & \null & \null & \null & \null & \null & \null \\
       \so(\mathrm{m},\mathrm{m}) \, \supset \, \mathfrak{L} & = & \left (\begin{array}{cc}
\mathfrak{A} & \mathfrak{B} \\
\mathfrak{C} & \mathfrak{D}
\end{array}\right ) & \Rightarrow & \mathfrak{L}^T \, \mathbb{C}_{\mathrm{off}} \, + \, \mathbb{C}_{\mathrm{off}} \mathfrak{L} & = & 0
     \end{array}
\end{equation}
By explicit evaluation of the Lie algebra conditions we find:
\begin{equation}\label{liecondo}
    \mathfrak{D} \, = \, - \, \mathfrak{A}^T \quad ; \quad \mathfrak{B} \, = \, - \, \mathfrak{B}^T \quad ; \quad \mathfrak{C} \, = \, - \, \mathfrak{C}^T
\end{equation}
Let us now consider the image of the Lie algebra element $\mathfrak{L}$ in the diagonal-basis:
\begin{eqnarray}
{\cal S} & \equiv & {\cal C}_{\mathrm{so}}\, \mathfrak{L} \, {\cal C}_{\mathrm{so}} \, = \, {\frac{1}{2}} \left(
\begin{array}{cc}
 \left ( \mathfrak{A} + \mathfrak{B}  + \mathfrak{C} + \mathfrak{D} \right) &
 \left ( \mathfrak{A} -  \mathfrak{B}  + \mathfrak{C}  - \mathfrak{D} \right) \\
\left ( \mathfrak{A} + \mathfrak{B}  - \mathfrak{C} -  \mathfrak{D} \right) &
\left ( \mathfrak{A} -  \mathfrak{B}  - \mathfrak{C}  + \mathfrak{D} \right)
\end{array}
\right)
\label{mappetta2}
\end{eqnarray}
If we impose the conditions $\mathfrak{A}=\mathfrak{D}$ and $\mathfrak{B}=\mathfrak{C}$
we obtain the subalgebra $\so(\mathrm{m}) \times \so(\mathrm{m}) \subset \so(\mathrm{m, m})$:
\begin{equation}
 \so(\mathrm{m}) \times \so(\mathrm{m}) \,\ni\,\left ( \begin{array}{cc} \mathfrak{A} + \mathfrak{B} &0  \cr 0 & \mathfrak{A}-\mathfrak{B}
 \end{array} \right )\quad ; \quad \mathfrak{A} \, = \, - \mathfrak{A}^T \quad ; \quad \mathfrak{B} \, = \, - \, \mathfrak{B}^{T}
\end{equation}
The basic strategy to obtain the general formula for the kinetic matrix $\mathcal{M}$ of $\mathrm{m}$-twisted scalars coupled to a $\mathcal{G}/\mathcal{H}$ sigma-model  is completely analogous to that employed in $D=4$ theories. There must exist a pseudo-orthogonal embedding
\begin{equation}
\boxed{
  \mathcal{G} \mapsto \mathrm{SO(m,m) }
\label{ortoembed}}
\end{equation}
 such that the
isotropy subgroup ${\cal H}\subset {\cal G}$ gets embedded into the
maximal compact subgroup $\mathrm{SO}(m)\times \mathrm{SO}(m)$, namely:
\begin{eqnarray}
{\cal G}  {\stackrel{\iota_\delta}{\longrightarrow}}  \, \mathrm{SO} (\mathrm{m,m})\,; \quad
{\cal H}
{\stackrel{\iota_\delta}{\longrightarrow}} \,  \mathrm{SO}(\mathrm{m}) \times \mathrm{SO}(\mathrm{m})
\label{gruppino2}
\end{eqnarray}
Relying on the orthogonal embedding of eq.~(\ref{gruppino2})
we obtain the following pseudo-orhtogonal representation of the coset representative
$\mathbb{L}$:
\begin{eqnarray}\label{frescoletto}
    \mathbb{L}(\phi)  \,\longrightarrow \, \left ( \begin{array}{cc} \mathcal{A}(\phi) & \mathcal{B}(\phi)\cr
\mathcal{C}(\phi) & \mathcal{D}(\phi) \end{array} \right ) & \equiv & \mathcal{O}(\phi) \, \in \, \mathrm{SO(m,m)}\nonumber\\
&& \mathcal{O}^T (\phi)\, \mathbb{C}_{\mathrm{off}} \, \mathcal{O}(\phi) \, = \, \mathbb{C}_{\mathrm{off}}
\end{eqnarray}
Next in full analogy with equation (\ref{mixarepra1}) let us introduce the mixed-basis representation
of the coset representative
\begin{eqnarray}
\mathcal{O}_{\mathrm{mr}}(\phi)&\equiv &\left(\begin{matrix}\mathbf{ f}(\phi)
&{\mathbf{ \widetilde f}}(\phi)\cr \mathbf{ h}(\phi) &{\mathbf{\widetilde h}}(\phi)
\end{matrix}\right)=\mathcal{O}(\phi)\,\mathcal{C}_{\mathrm{so}}\,,\label{mixarepra2}
\end{eqnarray}
where, by explicit evaluation we have:
\begin{equation}
\begin{array}{ccccccc}
  \mathbf{ f}(\phi) &=& \frac{1}{\sqrt{2}}\left (\mathcal{A} + \mathcal{B}\right) & ; & \tilde{\mathbf{ f}}(\phi) &=& \frac{1}{\sqrt{2}}\left (\mathcal{A} - \mathcal{B}\right) \\
  \mathbf{ h}(\phi) &=& \frac{1}{\sqrt{2}}\left (\mathcal{C} + \mathcal{D}\right) & ; & \tilde{\mathbf{ h}}(\phi) &=& \frac{1}{\sqrt{2}}\left (\mathcal{C} - \mathcal{D}\right) \\
  \end{array}
\end{equation}
and, from the pseudo-orthogonal relations imposed on the $\mathcal{A},\mathcal{B},\mathcal{C},\mathcal{D}$ blocks we deduce the following relations:
\begin{equation}\label{fhreleSO}
    \mathbf{f}^T \, \mathbf{h} \, + \, \mathbf{h}^T \, \mathbf{f} \, = \, \mathbf{1} \quad ; \quad \tilde{\mathbf{f}}^T \, \mathbf{h} \, + \, \tilde{\mathbf{h}}^T \, \mathbf{f} \, = \, 0
\end{equation}
that are the pseudo-orthogonal counterpart of the symplectic relations (\ref{fhrels}).
Then in full analogy with the symplectic $D=4$ case the period matrix $\mathcal{M}(\phi)$ can be defined as:
\begin{eqnarray}
\mathcal{M}(\phi)&\equiv
&\mathbf{h}(\phi)\,\mathbf{f}(\phi)^{-1}=
\left[\mathcal{C}(\phi) + \mathcal{D}(\phi)\right] \left[ \mathcal{A}(\phi) + \mathcal{B}(\phi)\right]^{-1}\,\label{masterformula2}
\end{eqnarray}
By the same token as in the previous case, the period matrix $\mathcal{M}(\phi)$ given by the generalized Gaillard-Zumino formula (\ref{masterformula2}) transforms correctly under the action of the group $\mathcal{G}$, namely as in eq.s(\ref{Mtransfa2}).
\section{Electric/magnetic superdualities in $D=2$ Bose-Fermi field
theories}\label{sec2}
We can now consider the generalization of the $D=2$ action (\ref{d2generlag}) to the case where the fields are both
of Bose and Fermi type. It has the following form (we have choosen, without loosing in generality,
the conformal gauge for the worldsheet metric):
\begin{eqnarray}
S_{(D=2)} & = & \int \, d^2x \,\left\{- \ft 1 2 \, H_{AB}(\Phi) \partial_\mu \Phi^A \partial ^\mu \Phi^B
\right. \nonumber\\
\null & \null & \left. +   \, \ft 12 \,\kappa \, \left [-
\partial_\mu \Pi^\Sigma \,  \Gamma_{\Sigma\Lambda}(\Phi) \, \partial^\mu \Pi^\Lambda \,
+ \partial_\mu \Pi^\Sigma \, \Theta_{\Sigma\Lambda}(\Phi)  \, \partial_\nu \Pi^\Lambda \, \epsilon^{\mu\nu} \right]\right\}
\label{d2generlag-super}
\end{eqnarray}
where $\Phi$ are super-coordinates parameterizing
a supermanifold $\mathcal{SM}^{(x|y)}$, which can be, in particular, a supercoset $\widehat{\mathcal{G}}/\widehat{\mathcal{H}}$, while $\Pi^\Sigma$ are a set of $(m|2n)$ fields of which the former $m$ are bosonic, while the latter $2n$ are fermionic.
The notation is as follows:
\begin{eqnarray}\label{indexconvention}
  A &=& \left\{ \underbrace{a}_{1,\dots,x} \, , \, \underbrace{\bar{a}}_{\bar{1},\dots,\bar{y}} \right\}\\
  \Lambda &=& \left\{ \underbrace{\alpha}_{1,\dots,m}\, , \, \underbrace{\overline{\alpha}}_{\bar{1},\dots,{\overline{2n}}} \right\}
\end{eqnarray}
where we have used the convention that unbarred indices are bosonic, while barred ones are fermionic. Furthermore $x$ is the bosonic dimension of the supermanifold $\mathcal{SM}^{(x|y)}$, while $y$ denotes its fermionic dimension.
Let us also note that, taking into account the statistics of the fields, we have the following graded symmetry and graded antisymmetry of the super-matrices entering action (\ref{d2generlag-super}):
\begin{equation}\label{grasimma}
\begin{array}{ccccc}
    H_{AB}(\Phi) & = & (-)^{AB} \, H_{BA}(\Phi) & \Rightarrow & \left \{ \begin{array}{ccc}
                                                                           H_{ab} & = & H_{ba} \\
                                                                           H_{\bar{a}\bar{b}} & = & - \, H_{\bar{b}\bar{a}}  \\
                                                                           H_{a\bar{b}} & = & H_{\bar{b}{a}}
                                                                         \end{array}
    \right.\\
     \Gamma_{\Lambda\Sigma}(\Phi) & = & (-)^{\Lambda\Sigma} \, \Gamma_{\Sigma\Lambda}(\Phi) &\Rightarrow &\left \{ \begin{array}{ccc}
                                                                           \Gamma_{\alpha\beta} & = & \Gamma_{\beta\alpha} \\
                                                                           \Gamma_{\bar{\alpha}\bar{\beta}} & = & - \, \Gamma_{\bar{\beta}\bar{\alpha}}  \\
                                                                           \Gamma_{\alpha\bar{\beta}} & = & \Gamma_{\bar{\beta}{\alpha}}
                                                                         \end{array}
    \right.\\
    \Theta_{\Lambda\Sigma}(\Phi) & = & - (-)^{\Lambda\Sigma} \, \Theta_{\Sigma\Lambda}(\Phi) &\Rightarrow &\left \{ \begin{array}{ccc}
                                                                           \Theta_{\alpha\beta} & = & - \,\Theta_{\beta\alpha} \\
                                                                           \Theta_{\bar{\alpha}\bar{\beta}} & = &  \, \Theta_{\bar{\beta}\bar{\alpha}}  \\
                                                                           \Theta_{\alpha\bar{\beta}} & = & \, - \, \Theta_{\bar{\beta}{\alpha}}
                                                                         \end{array}
    \right.\\
    \end{array}
\end{equation}
\par
Let us  now consider the supergroup $\widehat{\mathcal{G}}$ of super-isometries of the $\sigma$-model super-metric
$H_{AB}(\Phi)$. In the case $\mathcal{SM}^{(x|y)}$ is a supercoset, $\widehat{\mathcal{G}}$ coincides with the numerator supergroup, yet what we are about to say has a wider range of validity:  It is not necessary that $\mathcal{SM}^{(x|y)}$
be a homogeneous supermanifold, it is sufficient for it to have some (super)-group of isometries $\widehat{\mathcal{G}}$ that can be continuous or even discrete.
\par In papers \cite{Beisert:2008iq,Berkovits:2008ic}, it has been shown that the GS action (or the corresponding pure spinor action) displays a similar structure on $AdS_5 \times S^5$ background. In that case, 4 bosonic coordinates (out of the 10 of the spacetime) and 8 fermionic coordinates can be taken as the fields $\Pi^\Lambda$ of our sigma model. The other coordinates enter the couplings $\Gamma$ and $\Theta$ and they can be view as the coordinates of the manifold $\mathcal{SM}^{(x|y)}$.

\par The important point to stress is that, in full analogy to purely bosonic theories the  symmetries of the $\Phi$ sector of the Lagrangian can be extended to a duality symmetry of the equations of motion and Bianchi identities of the complete theory (including also the $\Pi$.s) if and only if the following two conditions are satisfied:
\begin{description}
  \item[a] There exists an orthosymplectic embedding:
\begin{equation}\label{superemba}
    \widehat{\mathcal{G}} \, \mapsto \, \mathrm{\OSp(m,m|4n)}
\end{equation}
\item[b] The kinetic matrix:
\begin{equation}\label{superemme}
    \widehat{\mathcal{M}}(\Phi) \, \equiv \, \widehat{\Gamma}(\Phi) \, + \, \widehat{\Theta}(\Phi)
\end{equation}
is acted on by the $\mathrm{\OSp(m,m|4n)}$ realization of
$\widehat{\mathcal{G}}$ with suitable \textit{fractional linear
transformations}.
\end{description}
Let us prove the above statements in some detail.
\subsection{Orthosymplectic duality symmetries}
Let us define the required orthosymplectic embedding of the isometry supergroup.
Each element $\widehat{\xi} \, \in \, \widehat{\mathcal{G}}$ is mapped into a graded matrix:
\begin{equation}
  \widehat{\xi} \mapsto \widehat{\Lambda}_\xi \equiv \left( \begin{array}{c|c}
     \widehat{\mathcal{A}}_\xi & \widehat{\mathcal{B}}_\xi \\
     \hline
     \widehat{\mathcal{C}}_\xi & \widehat{\mathcal{D}}_\xi \
  \end{array} \right)
\label{embeddusmatra2}
\end{equation}
that satisfies the defining equation:
\begin{eqnarray}
  \widehat{\Lambda}_\xi ^T \, \widehat{\mathbb{C}}_{\mathrm{off}} \, \widehat{\Lambda}_\xi =  \widehat{\mathbb{C}}_{\mathrm{off}}\,,\quad
 \widehat{\mathbb{C}}_{\mathrm{off}}  =   \left( \begin{array}{c|c}
     \mathbf{0}_{(m + 2n) \times (m + 2n)}  &  \mathbf{\Omega}_{(m + 2n) \times (m + 2n)} \\
     \hline
      \mathbf{\Omega}_{(m + 2n) \times (m + 2n)}  & \mathbf{0}_{(m + 2n) \times (m + 2n)}  \
  \end{array} \right) \label{definingortosymp}
\end{eqnarray}
where $\Omega$ is the invariant metric for an
$\mathrm{\OSp}\mathrm{(m|2n)}$ superalgebra. For instance one can
choose:
\begin{eqnarray}
  \mathbf{\Omega}_{(m + 2n) \times (m + 2n)} & = & \left( \begin{array}{cc}
                                                            \eta_{m \times m} & 0_{m \times 2n} \\
                                                            0_{m \times 2n}  & \epsilon_{2n \times 2n}
                                                          \end{array}
  \right)\nonumber\\
  \eta^T & = & \eta \quad ; \quad \eta \,\eta \, = \, \mathbf{1} \quad ; \quad \mbox{signature} \, = \, (+)^m\,\,
  \Rightarrow \eta = {\bf 1}_m  \nonumber\\
  \epsilon^T & = & - \epsilon \quad ; \quad \epsilon \,\epsilon \, = \, - \, \mathbf{1}
\label{definingortosympOmega}
\end{eqnarray}
but there are also other choices discussed in the sequel.
\par
The above definition implies the following relations on the supermatrix blocks:
\begin{eqnarray}\label{relSuper1}
&&  \widehat{\mathcal{A}}_\xi^{ST}\, {\bf \Omega} \, \widehat{\mathcal{C}}_\xi
\,+ \, \widehat{\mathcal{C}}_\xi^{ST}\, {\bf \Omega} \widehat{\mathcal{A}}_\xi \, = \, 0\,
\hspace{1cm}  \widehat{\mathcal{A}}_\xi^{ST}\, {\bf \Omega} \, \widehat{\mathcal{D}}_\xi
\,+ \, \widehat{\mathcal{C}}_\xi^{ST}\, {\bf \Omega} \widehat{\mathcal{B}}_\xi \, = \, {\bf \Omega}
\nonumber\\
&&  \widehat{\mathcal{B}}_\xi^{ST}\, {\bf \Omega} \, \widehat{\mathcal{D}}_\xi
\,+ \, \widehat{\mathcal{D}}_\xi^{ST}\, {\bf \Omega} \widehat{\mathcal{B}}_\xi \, = \, 0\,
\hspace{1cm}  \widehat{\mathcal{B}}_\xi^{ST}\, {\bf \Omega} \, \widehat{\mathcal{C}}_\xi
\,+ \, \widehat{\mathcal{D}}_\xi^{ST}\, {\bf \Omega} \widehat{\mathcal{A}}_\xi \, = \, {\bf \Omega}.
\end{eqnarray}
where the superscript $ST$ denotes the usual conjugation for supermatrices.
Notice that the  matrix $\Omega$ satisfies the following important properties
\begin{equation}\label{STmatrix1}
{\Omega}^{ST} = {\Omega}
\left(
\begin{array}{ccc}
 {\bf 1}_m & {\bf 0}_{m\times 2n}  \\
{\bf 0}_{2n\times m}  &  -{\bf 1}_{2n}
\end{array}
\right) \quad ; \quad \Omega^2 \, = \, \left( \begin{array}{ccc}
 {\bf 1}_m & {\bf 0}_{m\times 2n}  \\
{\bf 0}_{2n\times m}  &  -{\bf 1}_{2n}
\end{array}
\right)
\end{equation}
which we assume in any case. In addition, we have $(\Omega^{ST})^{ST} = \Omega$ and $\Omega^{ST} = \Omega^3$. They will be very useful in establishing the form of the orthosymplectic duality symmetries. Indeed eq.(\ref{STmatrix1}) can be used to restate the symmetry properties (\ref{grasimma}) of the  coupling matrices $\Gamma$ and $\Theta$ appearing in the action (\ref{d2generlag-super}) in the following way:
\begin{equation}\label{HGammasymmetry}
    \widehat{\Gamma}^{ST} \, = \, \widehat{\Gamma} \, \Omega^2 \quad ; \quad \widehat{\Theta}^{ST} \, = \, - \, \widehat{\Theta} \, \Omega^2
\end{equation}
\par
The reader can check that the above formulae are formally analogous to
those for the purely bosonic case, the main difference being that
everyone wears a hat and is a graded matrix. Before proceeding some
comments are in order. Customarily graded matrices are written in
block-form so that the off-diagonal blocks are fermionic and the
diagonal ones are bosonic. This is not necessary. Graded matrices
can also be written in block-forms where each block is in turn a
graded matrix. This has been used in the above discussion. There
always exists a change of basis that reduces any such matrix
to the standard form where the fermionic entries
are grouped into off-diagonal blocks. In the usual description
the orthosymplectic group $\mathrm{\OSp(m,m|4n)}$ would be defined
as the set of graded matrices leaving the following \textit{metric} invariant:
\begin{equation}\label{Cdiaga1}
    \widehat{\mathbb{C}}^\prime_{\mathrm{dia}} \, = \, \left( \begin{array}{c|c||c|c}
                                                         \eta & 0 & 0 & 0 \\
                                                         \hline
                                                         0 & -\eta & 0 & 0 \\
                                                         \hline
                                                         \hline
                                                         0 & 0 & \epsilon & 0 \\
                                                         \hline
                                                         0 & 0 & 0 & -\epsilon
                                                       \end{array}
    \right )\,.
\end{equation}
By exchanging the second with the third row in the above matrix, we get:
\begin{eqnarray}\label{Cdiaga}
    \widehat{\mathbb{C}}_{\mathrm{dia}} \, = \,  \mathcal{P} \, \widehat{\mathbb{C}}^\prime_{\mathrm{dia}} \, \mathcal{P} =
    \left( \begin{array}{c|c||c|c}
                                                         \eta & 0 & 0 & 0 \\
                                                         \hline
                                                         0 & \epsilon & 0 & 0 \\
                                                         \hline
                                                         \hline
                                                         0 & 0 & -\eta & 0 \\
                                                         \hline
                                                         0 & 0 & 0 & -\epsilon
                                                       \end{array}
    \right )\,,\quad
    \mathcal{P}  =  \left(\begin{array}{c|c|c|c}
 1 & 0 & 0 & 0 \\
 \hline
 0 & 0 & 1 & 0 \\
 \hline
 0 & 1 & 0 & 0 \\
 \hline
 0 & 0 & 0 & 1
\end{array} \right)
\end{eqnarray}
It is also very simple to work out the transformation connecting the bases defined by $\widehat{\mathbb{C}}_{\mathrm{dia}}$ and by $\widehat{\mathbb{C}}_{\mathrm{off}}$, respectively. It is given by the following generalized \textit{super-Cayley matrix}:
\begin{equation}\label{Cambiatone}
  \mathcal{C}_{\mathrm{\OSp}} \, = \, \frac{1}{\sqrt{2}}\, \left(
\begin{array}{c||c|c||c}
 1 & 0 & 1 & 0 \\
 \hline
 \hline
 0 & 1 & 0 & 1 \\
 \hline
 1 & 0 & -1 & 0 \\
 \hline
 \hline
 0 & 1 & 0 & -1
\end{array}
\right)
\end{equation}
Indeed, we have:
\begin{equation}\label{lulalula}
    \mathcal{C}_{\mathrm{\OSp}} \, \widehat{\mathbb{C}}_{\mathrm{dia}} \,\mathcal{C}_{\mathrm{\OSp}}^{-1}  = \, \widehat{\mathbb{C}}_{\mathrm{off}}  \, \equiv \, \left( \begin{array}{c||c|c||c}
                                                         0 & 0 & \eta & 0 \\
                                                         \hline
                                                         \hline
                                                         0 & 0 & 0 & \epsilon \\
                                                         \hline
                                                         \eta & 0 & 0 & 0 \\
                                                         \hline
                                                         \hline
                                                         0 & \epsilon & 0 & 0
                                                       \end{array}
    \right )
\end{equation}
So we can easily go from one basis to the other by means of these transformations and we can define the orthosymplectic group as in eq. (\ref{definingortosymp}).
\par
This being clarified let us proceed with the discussion of the fractional linear transformations of the matrix $\widehat{\mathcal{M}}$.
To this effect consider the following situation. Suppose that we have two supermatrices $\widehat{\mathcal{X}}$ and $\widehat{\mathcal{Y}}$ which have the same linear fractional transformation under a supermatrix
$\left( \begin{array}{c|c}
\widehat{\mathcal{A}} & \widehat{\mathcal{B}} \\
\hline
\widehat{\mathcal{C}} & \widehat{\mathcal{D}}\
\end{array} \right)$:
\begin{eqnarray}
  \widehat{\mathcal{X}}^\prime &=& \left( \widehat{\mathcal{C}} + \widehat{\mathcal{D}} \widehat{\mathcal{X}} \right ) \, \left ( \widehat{\mathcal{A}} + \widehat{\mathcal{B}} \widehat{\mathcal{X}} \right)^{-1}\,,\label{Xtrasfa} \\
  \widehat{\mathcal{Y}}^\prime &=& \left( \widehat{\mathcal{C}} + \widehat{\mathcal{D}} \widehat{\mathcal{Y}} \right ) \, \left ( \widehat{\mathcal{A}} + \widehat{\mathcal{B}} \widehat{\mathcal{Y}} \right)^{-1}\,,\label{Ytrasfa}
\end{eqnarray}
and let us formulate the following question: Which linear transposition relation between $\widehat{\mathcal{X}}$ and $\widehat{\mathcal{Y}}$ will imply, as consistency
conditions, the orthosymplectic relations (\ref{relSuper1}) on the supermatrix blocks $\widehat{\mathcal{A}},\widehat{\mathcal{B}},\widehat{\mathcal{C}},\widehat{\mathcal{D}}$? Such a question is relevant for the issue of duality rotations since it is precisely in this way that the symplectic or pseudo-orthogonal character of the
duality transformations is established in bosonic theories, by comparing the action of the latter on self-dual and anti-self dual field strengths.
Eq.s (\ref{Xtrasfa})-(\ref{Ytrasfa}) are consistent with the orthosymplectic conditions (\ref{relSuper1})  if and only if:
\begin{equation}\label{relazionaXY}
    \widehat{\mathcal{Y}} \, = \, - \, \Omega \, \widehat{\mathcal{X}}^{ST} \, \Omega^{ST}\,.
\end{equation}
Indeed, by calculating the super-transposed of the transformed  $\widehat{\mathcal{Y}}$ we get:
\begin{eqnarray}\label{feconda1}
    \left(\widehat{\mathcal{Y}}^\prime\right)^{ST} = \, - \, \Omega \, \widehat{\mathcal{X}}^\prime \, \Omega^{ST} \,
     = \, - \, \left(\widehat{\mathcal{A}}^{ST} + \widehat{\mathcal{Y}}^{ST} \widehat{\mathcal{B}}^{ST} \right )^{-1} \, \left(\widehat{\mathcal{C}}^{ST} + \widehat{\mathcal{Y}}^{ST} \widehat{\mathcal{D}}^{ST} \right)
\end{eqnarray}
Combining (\ref{feconda1}) with the transformation law of $\widehat{\mathcal{X}}$ and using the property $\Omega^{ST}\,\Omega \, = \, \mathbf{1}$, we obtain the relation:
\begin{equation}\label{feconda2}
    \left (\widehat{\mathcal{C}} + \widehat{\mathcal{D}} \widehat{\mathcal{X}}\right)\,\left ( \widehat{\mathcal{A}} + \widehat{\mathcal{B}} \widehat{\mathcal{X}} \right)^{-1} \, = \, - \, \Omega^{ST} \, \left(\widehat{\mathcal{A}}^{ST} + \widehat{\mathcal{Y}}^{ST} \widehat{\mathcal{B}}^{ST} \right )^{-1} \, \left(\widehat{\mathcal{C}}^{ST} + \widehat{\mathcal{Y}}^{ST} \widehat{\mathcal{D}}^{ST} \right) \, \Omega
\end{equation}
Multiplying the above relation to the right by $\left ( \widehat{\mathcal{A}} + \widehat{\mathcal{B}} \widehat{\mathcal{X}} \right)\, \Omega$ and to the left by the following supermatrix $\left(\widehat{\mathcal{A}}^{ST} + \widehat{\mathcal{Y}}^{ST} \widehat{\mathcal{B}}^{ST} \right ) \, \Omega$, we obtain:
\begin{eqnarray}
  0 &=& \left( \widehat{\mathcal{A}}^{ST} \,\Omega\, \widehat{\mathcal{C}} + \widehat{\mathcal{C}}^{ST} \,\Omega\, \widehat{\mathcal{A}}\right) + \widehat{\mathcal{Y}}^{ST} \left( \widehat{\mathcal{D}}^{ST} \,\Omega\, \widehat{\mathcal{B}} + \widehat{\mathcal{B}}^{ST} \,\Omega\, \widehat{\mathcal{D}}\right)  \nonumber\\
  \null && + \left( \widehat{\mathcal{C}}^{ST} \,\Omega\, \widehat{\mathcal{B}} + \widehat{\mathcal{A}}^{ST} \,\Omega\, \widehat{\mathcal{D}}\right) \,\widehat{\mathcal{ X}} + \widehat{\mathcal{Y}}^{ST} \left( \widehat{\mathcal{D}}^{ST} \,\Omega\, \widehat{\mathcal{A}} + \widehat{\mathcal{B}}^{ST} \,\Omega\, \widehat{\mathcal{C}}\right) \widehat{\mathcal{ X}}
\end{eqnarray}
which is satisfied given the relation (\ref{relazionaXY}) if and only if the orthosymplectic conditions (\ref{relSuper1})
are fulfilled by the $\widehat{\mathcal{A}},\widehat{\mathcal{B}},\widehat{\mathcal {C}},\widehat{\mathcal {D}}$ blocks.
\par
Let us now consider the following case of matrix $\widehat{\mathcal{X}}$:
\begin{equation}
\label{Xmatra}
    \widehat{\mathcal{X}} \, = \, \Omega \, \widehat{\mathcal{M}} \, = \, \Omega \, \left (\widehat{\Gamma} + \widehat{\Theta} \right)
\end{equation}
The corresponding $\widehat{\mathcal{Y}}$ has the following expression:
\begin{equation}
\label{Ymatra}
    \widehat{\mathcal{Y}} \, = \, - \, \Omega \, \widehat{\mathcal{M}}^{ST} \, \Omega^2 \, = \, - \Omega \left(\widehat{\Gamma} \, - \, \widehat{\Theta}\right)
\end{equation}
the last equality following from eq.(\ref{HGammasymmetry}).
\par
Consider now the action (\ref{d2generlag-super}) and in analogy with the bosonic case let us define the 1-form field strengths
of the twisted Bose/Fermi scalars as follows:
\begin{eqnarray}
F^{\Lambda} & = & d\Pi^\Lambda \quad ; \quad F^{\Lambda\pm} \, = \, \ft 12 \left ( F^\Lambda \pm \null^\star F^{\Lambda} \right)
\nonumber\\
G_\Sigma & = & \ft 12 \, \frac{\delta \, \mathcal{L}}{\delta \null^\star F^{\Lambda}} ; \quad G^{\pm}_\Sigma \, = \, \ft 12 \left ( G_\Sigma \pm \null^\star G_\Sigma \right)
\nonumber
\label{fildeforze}
\end{eqnarray}
From the above definitions we immediately derive the following relations:
\begin{eqnarray}\label{relazioniGF}
    G^{+} & = & \, \null \, \,\,\left ( \widehat{\Gamma} + \widehat{\Theta}\right ) \, F^{+} \, = \, \null \,\,\widehat{\mathcal{M}} \, F^{+}\nonumber\\
    G^{-} & = & - \left ( \widehat{\Gamma} - \widehat{\Theta}\right) \, F^{-} \, = \, - \widehat{\mathcal{M}}^{ST} \, \Omega^2 \, F^{+}
\end{eqnarray}
We can assume the following realization of the $\widehat{\mathcal{G}}$ superisometries on the electric and magnetic field strengths, we find:
\begin{equation}\label{feritino}
    \forall \, \xi \, \in \, \widehat{\mathcal{G}} \, : \quad \iota_\xi \, \left (\begin{array}{c}
                                                                                 F^\pm \\
                                                                                 \Omega \,G^\pm
                                                                               \end{array}
     \right) \, = \, \left (\begin{array}{cc}
                              \widehat{\mathcal{A}}_\xi & \widehat{\mathcal{B}}_\xi \\
                              \widehat{\mathcal{C}}_\xi & \widehat{\mathcal{D}}_\xi
                            \end{array}
      \right) \, \left (\begin{array}{c}
                                                                                 F^\pm \\
                                                                                 \Omega \,G^\pm
                                                                               \end{array}
     \right)
\end{equation}
which are consistent with the relation (\ref{relazioniGF}) if the matrices $\widehat{\mathcal{X}}$ and $\widehat{\mathcal{Y}}$, as defined in eq.s (\ref{Xmatra}) and (\ref{Ymatra}), transform according to eq.(\ref{Xtrasfa})
and (\ref{Ytrasfa}) under $\left (\begin{array}{cc}
                              \widehat{\mathcal{A}}_\xi & \widehat{\mathcal{B}}_\xi \\
                              \widehat{\mathcal{C}}_\xi & \widehat{\mathcal{D}}_\xi
                            \end{array}
      \right)$.
This concludes the discussion of duality symmetries in general Bose/Fermi theories.
\subsection{The Gaillard-Zumino formula in the orthosymplectic case}
Having clarified the general form of the duality covariant Bose-Fermi sigma--models, we can now focus on the case where the
supermanifold  $\mathcal{SM}^{(x|y)}\, \equiv \, \widehat{\mathcal{G}}/\widehat{\mathcal{H}}$ is a homogeneous super-coset.
In this case we can easily construct the orthosymplectic generalization of the Gaillard-Zumino formula.
What is important to notice is that the basis $\widehat{\mathbb{C}}_{\mathrm{dia}}$ is the one in which the subalgebra
\begin{equation}\label{diagaosp}
    \osp(\mathrm{m|2n}) \times \osp(\mathrm{m|2n}) \, \subset \, \osp(\mathrm{m,m|4n})
\end{equation}
is diagonally embedded:
\begin{equation}\label{diagapussa}
    \left( \begin{array}{cc}
             \osp(\mathrm{m|2n})_I& 0 \\
             0 & \osp(\mathrm{m|2n})_{II}
           \end{array}
    \right ) \, \in \, \osp(\mathrm{m,m|4n})
\end{equation}
Correspondingly, given a coset representative $\widehat{\mathbb{L}}(\Phi)$ of $\widehat{\mathcal{G}}/\widehat{\mathcal{H}}$,
we can consider its orthosymplectic embedding in the off-diagonal basis:
\begin{eqnarray}\label{frescolettoSp}
    \widehat{\mathbb{L}}(\Phi)  \,\longrightarrow \, \left ( \begin{array}{cc} \widehat{\mathcal{A}}(\Phi) & \widehat{\mathcal{B}}(\Phi)\cr
\widehat{\mathcal{C}}(\Phi) & \widehat{\mathcal{D}}(\Phi) \end{array} \right ) & \equiv & \widehat{\mathcal{O}}(\Phi) \, \in \, \mathrm{\OSp(m,m|4n)}\,,\nonumber\\
\widehat{\mathcal{O}}^T (\Phi)\, \widehat{\mathbb{C}}_{\mathrm{off}} \, \widehat{\mathcal{O}}(\Phi) \, &=& \, \widehat{\mathbb{C}}_{\mathrm{off}}
\end{eqnarray}
As before, we also  require that the embedding of $\widehat{\mathcal{G}}$ is such that its subgroup $\widehat{\mathcal{H}}$ gets embedded into $\osp(\mathrm{m|2n}) \times \osp(\mathrm{m|2n}) \, \subset \, \osp(\mathrm{m,m|4n})$. Then
in full analogy with equation (\ref{mixarepra2}) let us introduce the mixed-basis representation
of the coset representative.
\begin{eqnarray}
\widehat{\mathcal{O}}_{\mathrm{mr}}(\Phi)&\equiv &\left(\begin{matrix}\widehat{\mathbf{ f}}(\Phi)
&\widehat{\widetilde{\mathbf{ f}}}(\Phi)\cr \widehat{\mathbf{ h}}(\Phi) &\widehat{\widetilde{\mathbf{
h}}}(\Phi)
\end{matrix}\right)=\widehat{\mathcal{O}}(\Phi)\,\widehat{\mathcal{C}}_{\mathrm{so}}^{-1}\,,\label{mixarepra3}
\end{eqnarray}
where, by explicit evaluation we have:
\begin{equation}
\begin{array}{ccccccc}
  \widehat{\mathbf{ f}}(\Phi) &=& \frac{1}{\sqrt{2}}\left (\widehat{\mathcal{A}}(\Phi) + \widehat{\mathcal{B}}(\Phi)\right) & ; & \widehat{\widetilde{\mathbf{ f}}}(\Phi) &=& \frac{1}{\sqrt{2}}\left (\widehat{\mathcal{A}}(\Phi) - \widehat{\mathcal{B}}(\Phi)\right) \\
  \widehat{\mathbf{ h}}(\Phi) &=& \frac{1}{\sqrt{2}}\left (\widehat{\mathcal{C}}(\Phi) + \widehat{\mathcal{D}}(\Phi)\right) & ; & \widehat{\widetilde{\mathbf{ h}}}(\Phi) &=& \frac{1}{\sqrt{2}}\left (\widehat{\mathcal{C}}(\Phi) - \widehat{\mathcal{D}}(\Phi)\right) \\
  \end{array}
\end{equation}
From the orthosymplectic relations imposed on the $\widehat{\mathcal{A}},\widehat{\mathcal{B}},\widehat{\mathcal{C}},\widehat{\mathcal{D}}$ blocks we deduce the following relations:
\begin{equation}\label{fhreleOSO}
    \widehat{\mathbf{f}}^{ST} \, \Omega \, \widehat{\mathbf{h}} \, + \, \widehat{\mathbf{h}}^{ST} \,\Omega \, \widehat{\mathbf{f}} \, = \, \Omega \quad ; \quad \widehat{\widetilde{\mathbf{f}}}{}^{ST} \, \Omega \, \widehat{\mathbf{h}} \, + \, \widehat{\widetilde{\mathbf{h}}}{}^{ST}\, \Omega \, \widehat{\mathbf{f}} \, = \, 0
\end{equation}
that are the orthosymplectic counterpart of the symplectic and pseudorthogonal relations (\ref{fhrels})-(\ref{fhreleSO}).
\par
Then the matrix $\widehat{\mathcal{X}} \, \equiv \, \Omega \widehat{\mathcal{M}}$ is defined by the obvious generalization of eq.(\ref{masterformula2})
\begin{equation}
\boxed{
\Omega \, \widehat{\mathcal{M}}(\Phi) \equiv
\widehat{\mathbf{h}}(\Phi)\,\widehat{\mathbf{f}}(\Phi)^{-1}=
\left[\widehat{\mathcal{C}}(\Phi) + \widehat{\mathcal{D}}(\Phi)\right] \left[ \widehat{\mathcal{A}}(\Phi) + \widehat{\mathcal{B}}(\Phi)\right]^{-1}}
\label{masterformula3}
\end{equation}
and by the same token as in the previous cases it  transforms correctly under the action of the supergroup $\widehat{\mathcal{G}}$.


\section{${\bf \mathrm{SO(m,m)}}$ embeddings from dimensional reduction $4D\, \rightarrow\,
2D$}\label{sec3} Let us now come back to the purely bosonic theories
and recall a phenomenon which was discovered in the context of
dimensional reduction and provides a challenging suggestion also for
Bose/Fermi theories.
\par
As we stressed in section \ref{d4sugrasym}, as long as there is no
gauging, the general form of the bosonic part of a four-dimensional
supergravity Lagrangian is given by eq.(\ref{d4generlag}). Let us
name $\mathrm{U_{4D}}$ the duality group which is an isometry of the
sigma-model part of that Lagrangian and acts by symplectic duality
symmetries on the vector fields. Then, following the discussion of
\cite{Fre':2005si}, we can perform the dimensional reduction $D=4
\rightarrow D=2$. There are two possible routes one can follow
\cite{d2}:
\begin{description}
\item[Ehlers:] The Ehlers route consists of two steps. In the first step one performs the dimensional reduction  to $D=3$ and then dualizes all-vector fields to scalars. In the second step one goes down to $D=2$. Following this route one arrives at a standard sigma model of the following form:
      \begin{equation}\label{elerone}
        {S}_{Ehlers}\, = \, \int \, d^2x \, \mathfrak{h}_{IJ} (\Upsilon) \, \partial_\mu \Upsilon^I\, \partial^\mu \Upsilon^J
      \end{equation}
      where $\mathfrak{h}_{IJ} (\Upsilon)$ is the invariant metric of a non-compact coset manifold $\mathbb{U_{3D}}/\mathrm{H_{3D}}$. The four dimensional duality group is a subgroup of the Ehlers group $\mathbb{U_{4D}} \subset \mathrm{U_{3D}}$ and we have the following general Lie algebra decomposition:
      \begin{equation}
\mbox{adj}(\mathbb{U}_{\mathrm{3D}}) =
\mbox{adj}(\mathbb{U}_{\mathrm{4D}})\oplus\mbox{adj}(\mathrm{SL(2,\mathbb{R})})\oplus
W_{(\mathbf{W}),2}
\label{gendecompo}
\end{equation}
where $\mathbf{W}$ is the {symplectic} representation of
$\mathbb{U}_{4D}$ to which the electric and magnetic field
strengths of the vector fields are assigned in order to construct
the duality symmetries of the four-dimensional Lagrangian
(\ref{d4generlag}).
      \item[Matzner-Missner:] The second dimensional reduction route, named after Matzner--Missner consists of stepping down directly from $D=4$ to $D=2$, where the scalars remain scalars and the vector fields yield 0-form scalars.
      The final action is of the form (\ref{d2generlag}). More precisely it is the following:
          \begin{eqnarray}
S^{MM}_{(D=2)} & = & \int \, d^2x \, \left \{
- \ft 1 2 \, h_{ab}(\phi) \partial_\mu \phi^a \partial ^\mu \phi^b
\right. \\
\null & + & \left.   \ft 12 \, \left [-
\nabla_\mu \pi^{\Lambda |A} \,  \mbox{Im} \mathcal{N}_{\Lambda\Sigma}(\phi) \, \delta_{AB}  \,
\nabla^\mu \pi^{\Sigma |B} \,
+ \nabla_\mu \pi^{\Lambda | A}  \, \mbox{Re} \mathcal{N}_{\Lambda\Sigma} \, \varepsilon_{AB}  \, \nabla_\nu
\pi^{\Sigma |B}
\, \epsilon^{\mu\nu} \right] \right \} \label{d2generlagMM} \nonumber
\end{eqnarray}
where $h_{ab}(\phi)$ is the invariant metric of the coset  $\mathbb{U}_{4D}/\mathrm{H_{4D}} \times \mathrm{SL(2,\mathbb{R})}/\mathrm{SO(2)}$, the second factor coming from the dimensional reduction of Einstein Gravity.
The symmetry $\mathbb{U}_{4D} \times  \mathrm{SL(2,\mathbb{R})}$ is realized by isometries on the sigma-model part of the Lagrangian and as duality transformations on 0-form-scalars through an embedding in $\mathrm{SO(m,m)}$ which we recall below.
\end{description}
Let $n$ be the number of vector fields appearing in  the $D=4$
supergravity action (\ref{d4generlag}). Then the dimension of the
symplectic representation $W$ of $\mathbb{U}_{4D}$ appearing in
eq.(\ref{gendecompo}) is $2n$ and the number of twisted scalars
appearing in the Matzner-Missner Lagrangian (\ref{d2generlagMM})
is also $2n$. Correspondingly the duality group is
$\mathrm{SO(2n,2n)}$ and the embedding
\begin{equation}\label{alettoinsieme}
    \mathrm{Sp(2n,\mathbb{R})} \, \mapsto \, \mathrm{SO(2n,2n)}
\end{equation}
was described in \cite{Fre':2005si}. As follows  $\forall \xi \in \mathbb{U}_{4D}$ let
$\left( \begin{array}{cc}
A_\xi & B_\xi \\
C_\xi & D_\xi
\end{array}
\right) \in \mathrm{Sp(2n,\mathbb{R})}$ be its representation by means of symplectic matrices that satisfy the defining condition:
\begin{equation}\label{deficondo1}
    \left( \begin{array}{cc}
A_\xi & B_\xi \\
C_\xi & D_\xi
\end{array}
\right)^T \, \left (\begin{array}{cc}
                      {0} & \mathbf{1} \\
                      -\mathbf{1} & {0}
                    \end{array}
 \right) \, \left( \begin{array}{cc}
A_\xi & B_\xi \\
C_\xi & D_\xi
\end{array}
\right) \, = \, \left (\begin{array}{cc}
                      {0} & \mathbf{1} \\
                      -\mathbf{1} & {0}
                    \end{array}
 \right)
\end{equation}
Then we have:
\begin{equation}\label{sottocoperta}
    \left( \begin{array}{cc}
A_\xi & B_\xi \\
C_\xi & D_\xi
\end{array}
\right) \, \mapsto \, \mathcal{O}_\xi \, \equiv \, \left (\begin{array}{c|c}
                               A_\xi \otimes \mathbf{1}_2 & B_\xi \otimes \epsilon \\
                               \hline
                               C_\xi \otimes \epsilon & A_\xi \otimes \mathbf{1}_2
                             \end{array}
 \right) \, \in \, \mathrm{SO(2n,2n)}
\end{equation}
where $\mathbf{1}_2$ is the identity matrix in two-dimension and $\epsilon \, = \, -\epsilon^T $ denotes an antisymmetric
$2\times 2$ matrix such that $\epsilon^2 = -\mathbf{1}_2$.

The constructed matrix $\mathcal{O}_\xi$ satisfies the pseudo-orthogonality conditions in the form:
\begin{equation}\label{pseudovecchia}
    \mathcal{O}_\xi^T \,
    \left (\begin{array}{c|c}
                               0 & \mathbf{1}_{n} \otimes \mathbf{1}_{2 } \\
                               \hline
                               \mathbf{1}_{n} \otimes \mathbf{1}_{2 }  & 0\
                               \end{array}
 \right) \, \mathcal{O}_\xi \, = \, \left (\begin{array}{c|c}
                               0 & \mathbf{1}_{n} \otimes \mathbf{1}_{2 } \\
                               \hline
                               \mathbf{1}_{n} \otimes \mathbf{1}_{2 }  & 0\
                               \end{array}
 \right)
\end{equation}
Correspondingly the period matrix $\mathcal{M}$ in two dimensions is related to the period matrix $\mathcal{N}$ in four dimensions through the following formula:
\begin{eqnarray}
     \mathcal{M }  & = & \mbox{Im} \mathcal{N} \otimes \mathbf{1}_{2}
     -\mbox{Re} \mathcal{N} \otimes \mathbf{\epsilon}
\end{eqnarray}
which yield the result displayed in the Lagrangian
(\ref{d2generlagMM}), namely:
\begin{eqnarray}
  (\gamma_{\alpha\beta}) & = & \mbox{Im} \mathcal{N} \otimes \mathbf{1}_{2}\, = \,
  (\mbox{Im} \mathcal{N}_{\Lambda\Sigma} \, \delta_{AB})
  \nonumber\\
 (\theta_{\alpha\beta}) & = & - \mbox{Re} \mathcal{N} \otimes \mathbf{1}_{2}\, = - \,
 (\mbox{Re} \mathcal{N}_{\Lambda\Sigma} \, \epsilon_{AB})
  \nonumber
\label{invasiocamp}
\end{eqnarray}
The main relevant point is that the Matzner-Missner
(\ref{d2generlagMM}) and the Ehlers Lagrangian (\ref{elerone}) can
be mapped into one another by a suitable duality transformation in
$\mathrm{SO(2n,2n)}/\mathbb{U}_{4D}$.
\par
In order to imitate the same mechanism of duality at the level of Bose/Fermi theories it is convenient
to rewrite the embedding (\ref{alettoinsieme}) suitable for generalization to orthosymplectic groups.
This is easily done by changing the basis both of the symplectic group and of the pseudo-orthogonal one.
Suppose that $n$ be even and observe that
\begin{eqnarray}
  \left( \begin{array}{c|c}
           {\bf 1} & 0 \\
           \hline
           0 & \eta_{2n}
         \end{array}
  \right) \,\left( \begin{array}{c|c}
                     0 & \mathbf{1} \\
                     \hline
                     \mathbf{1} & 0
                   \end{array}
  \right ) \, \left( \begin{array}{c|c}
           {\bf 1} & 0 \\
           \hline
           0 & \eta_{2n}
         \end{array}\right) &=& \left( \begin{array}{c|c}
                     0 & \eta_{2n} \\
                     \hline
                     \eta_{2n} & 0
                   \end{array}\right) \\
\left( \begin{array}{c|c}
           {\bf 1} & 0 \\
           \hline
           0 & -\epsilon_{n}
         \end{array}
  \right) \,\left( \begin{array}{c|c}
                     0 & \mathbf{1} \\
                     \hline
                     -\mathbf{1} & 0
                   \end{array}
  \right ) \, \left( \begin{array}{c|c}
           {\bf 1} & 0 \\
           \hline
           0 & \epsilon_{n}
         \end{array}\right) &=& \left( \begin{array}{c|c}
                     0 & \epsilon_{n} \\
                     \hline
                     \epsilon_{n} & 0
                   \end{array}\right)
\end{eqnarray}
where $\epsilon_{n}^T \, = \, - \epsilon_{n}$ is an antisymmetric matrix such that $\epsilon_{n}^2 \, = \, -\mathbf{1}$ and
$\eta^T_{2n} \, = \, \eta_{2n}$ is a symmetric one such that $\eta^2_{2n} \, = \, \mathbf{1}$. Hence by means of the transformation:
\begin{equation}\label{simpatrasfa}
    \left( \begin{array}{c|c}
           {\bf 1} & 0 \\
           \hline
           0 & -\epsilon_{n}
         \end{array}
  \right) \,\left( \begin{array}{c|c}
                     A_\xi & B_\xi \\
                     \hline
                     C_\xi & D_\xi
                   \end{array}
  \right ) \, \left( \begin{array}{c|c}
           {\bf 1} & 0 \\
           \hline
           0 & \epsilon_{n}
         \end{array}\right) \, = \, \left( \begin{array}{c|c}
                     X_\xi & Y_\xi \\
                     \hline
                     W_\xi & Z_\xi
                   \end{array}
  \right )  \, = \, \mathfrak{S}_\xi
\end{equation}
we obtain a symplectic matrix which satisfies the relations:
\begin{equation}\label{newformasimpa}
    \mathfrak{S}_\xi^T \, \left( \begin{array}{c|c}
                     0 & \epsilon_{n} \\
                     \hline
                     \epsilon_{n} & 0
                   \end{array}\right) \,\mathfrak{S}_\xi \, = \, \left( \begin{array}{c|c}
                     0 & \epsilon_{n} \\
                     \hline
                     \epsilon_{n} & 0
                   \end{array}\right)
\end{equation}
rather than in the form (\ref{deficondo1}). Similarly starting from any matrix $ \mathcal{O}$ which satisfies the pseudo-ortogonality conditions on the form (\ref{pseudovecchia}), by applying the following transformation:
\begin{equation}\label{fischio}
    \left( \begin{array}{c|c}
           {\bf 1} & 0 \\
           \hline
           0 & \eta_{2n}
         \end{array}
  \right) \,\mathcal{O}\, \left( \begin{array}{c|c}
           {\bf 1} & 0 \\
           \hline
           0 & \eta_{2n}
         \end{array}\right) \,  \equiv \, \mathfrak{O }
\end{equation}
we obtain a new one which satisfies them in the form:
\begin{equation}\label{pseudonuova}
    \mathfrak{O }^T \, \left( \begin{array}{c|c}
                     0 & \eta_{2n} \\
                     \hline
                     \eta_{2n} & 0
                   \end{array}\right)  \, \mathfrak{O } \, = \, \left( \begin{array}{c|c}
                     0 & \eta_{2n} \\
                     \hline
                     \eta_{2n} & 0
                   \end{array}\right)
\end{equation}
Consider now the case where $\eta_{2n}\, = \, \epsilon_n \otimes \epsilon_2$ and apply the transformation (\ref{fischio})
to the result of the embedding $\mathrm{Sp(2n,R)} \mapsto \mathrm{SO(2n,2n)}$ namely to the matrix $\mathcal{O}_\xi$
in eq.(\ref{sottocoperta}). By direct calculation we find:
\begin{equation}\label{furtiva}
    \mathfrak{O }_\xi \, = \, \left(\begin{array}{c|c}
                                      X\otimes \mathbf{1}_2 & Y\otimes \mathbf{1}_2 \\
                                      \hline
                                      W\otimes \mathbf{1}_2 & W\otimes \mathbf{1}_2
                                    \end{array}
     \right)
\end{equation}
This result  shows that the embedding of the
product group $
\mathrm{Sp(2n,\mathbb{R})}\times \mathrm{SL(2,\mathbb{R})} $ into $\mathrm{SO(2n,2n)}$, by changing
basis is just an instance of a general embedding
$\Sp(2p,\,\mathbb{R})\times \Sp(2q,\,\mathbb{R})\mapsto
\SO(2pq,\,2pq)$ defined, for even $p$, as follows:
\begin{eqnarray}
\mathrm{Sp(2p,\mathbb{R})} \ni \left( \begin{array}{c|c}
                     X_\xi & Y_\xi \\
                     \hline
                     W_\xi & Z_\xi
                   \end{array}
  \right )  & \rightarrow & \left(\begin{array}{c|c}
                                      X\otimes \mathbf{1}_{2q} & Y\otimes \mathbf{1}_{2q} \\
                                      \hline
                                      W\otimes \mathbf{1}_{2q} & W\otimes \mathbf{1}_{2q}
                                    \end{array}
     \right) \, \in \, \mathrm{SO(2pq,2pq)} \\
  \mathrm{Sp(2q,\mathbb{R})} \ni \Lambda & \rightarrow & \left(\begin{array}{c|c}
                                       \mathbf{1}_p\otimes \Lambda& 0 \\
                                      \hline
                                      0 & \mathbf{1}_p\otimes \Lambda
                                    \end{array}
     \right) \, \in \, \mathrm{SO(2pq,2pq)}
     \label{nuovoletto}
\end{eqnarray}
where the $\mathrm{SO(2pq,2pq)}$ invariant metric is:
$    \left( \begin{array}{c|c}
             0 & \epsilon_p\otimes \epsilon_{2q} \\
             \hline
             \epsilon_p \otimes \epsilon_{2q} & 0
           \end{array}
    \right)
$
and $\Lambda^T\,\epsilon_{2q} \, \Lambda \, = \, \epsilon_{2q}$ while $\left( \begin{array}{c|c}
                     X_\xi & Y_\xi \\
                     \hline
                     W_\xi & Z_\xi
                   \end{array}
  \right )$ satisfies eq.(\ref{newformasimpa}).
  \par
  Formulated as in eq.(\ref{nuovoletto}), the embedding can be extended to the orthosymplectic case
  giving rise to new Bose/Fermi analogues of the Ehlers/Matzner-Missner dual descriptions of the same
  physical system. This is what we study in the next section.
\section{Orthosymplectic $\OSp(m,m|4n)$ embeddings}\label{sec4}
Finally, we can merge the symplectic embeddings and the orthogonal ones.
In order to present the problem for the orthosymplectic emebeddings, we
proceed as follows. We first consider the direct product of two supergroups  and
a representation carrying an orthosymplectic structure. Then, we compute the
duality group needed to determine the GZ formula and the sigma model.
However, for reader's
convenience, we do it separately for the orthogonal subgroups as an example.
\par
We want to embed the product of
\begin{equation}\label{orpA}
\frac{\OSp(p,p|4r)}{\SO(p) \times \SO(p) \times {\rm U}(2r)} \times
\frac{\OSp(q,q|2s)}{\SO(q) \times \SO(q) \times {\rm U}(s)}
\end{equation}
into the coset
\begin{equation}\label{orpB}
\frac{\OSp(2pq + 4rs, 2 pq+ 4 rs| 4 ps + 8 qr)}{\SO(2pq + 4rs) \times \SO(2pq + 4rs) \times {\rm U}(2 ps + 4 qr)}
\end{equation}
whose super-isometry group acts on the representation ${ (4pq + 8rs | 4 ps + 8 qr)}$
with $4pq + 8rs$ bosons and $4 ps + 8 qr$ fermions. The embedding of the isometry group of (\ref{orpA}) into
the superisometry group of (\ref{orpB}) is such that the fundamental representation of the latter is the tensor product of the
fundamental representations of the two factor groups the former.

 The representation carries an orthosymplectic
structure and implements the duality relations for bosons, for fermions and their mixings.

Since the general result is rather cumbersome, we first exploit the embedding for the bosonic
subsectors. Now, we embed
\begin{equation}\label{orpBA}
\SO(p,p) \times \SO(q,q) \mapsto \SO(2pq,2pq)\,, \quad
{\rm Sp(4r)} \times {\rm Sp(2s)} \mapsto {\rm SO}(4 rs, 4rs)
\end{equation}
in order to obtain a linear representation of $\SO(n,n)$ with $n=2pq$ or $n=4rs$. Then, we
cast everything into a bigger representation of $\SO(2 pq + 4rs, 2pq+4 rs)$.

We first decompose $\Lambda_{2p} \in  \so(p,p) $ (we work at algebra level) into
 $$
\Lambda_{2p}
= \left(
\begin{array}{cc}
 A_p & B_p   \\
C_p & - A^T_p
 \end{array}
\right)\,, \quad B^T_p = - B_p\,, \quad C^T_p = - C_p\,.
 $$
which satisfies $\Lambda^T_{2p} \eta_{2p} + \eta_{2p} \Lambda_{2p} = 0$
where $$\eta_{2p} = \small
\left(
\begin{array}{cc}
 0 & {\bf 1}_p   \\
 {\bf 1}_p & 0
\end{array}
\right)$$
and we denote by $M_{2q}$ a matrix of $\so(q,q)$.
We get the new $4pq \times 4pq$ matrix of $\so(2pq, 2pq)$
 \begin{equation}\label{pipA}
\left(
\begin{array}{c|c}
A_p \otimes {\mathbf 1}_{2q} +  {\mathbf 1}_p \otimes M_{2q} &  B_p \otimes {\mathbf 1}_{2q} \\
\hline
C_p \otimes {\mathbf 1}_{2q} &  -  A^T_p \otimes {\mathbf 1}_{2q} -  {\mathbf 1}_p \otimes  M^T_{2q}
\end{array}
\right)
\end{equation}
where the invariant tensor is
 $$\eta_{4pq} = \small
\left(
\begin{array}{cc}
 0 & {\bf 1}_p \otimes {\eta}_{2q}   \\
 {\bf 1}_p \otimes {\eta}_{2q} & 0
\end{array}
\right) $$
In the same way,
we decompose $ \Lambda_{4r} \in  \sym(4r) $ into
 $$
 \Lambda_{2r}
= \left(
\begin{array}{cc}
  A_{2r} &  B_{2r}   \\
 C_{2r} &  \epsilon_{2r}  A^T_{2r} \epsilon_{2r}
 \end{array}
\right)\,, \quad  B^T_{2r} = \epsilon_{2r}  B_{2r} \epsilon_{2r}\,, \quad  C^T_{2r} = \epsilon_{2r}  C_{2r} \epsilon_{2r}\,.
 $$
which satisfies
$ \Lambda^T_{2r} \epsilon_{4r} + \epsilon_{4r}  \Lambda_{2r} = 0$
where
\begin{equation}\label{pifA}
\epsilon_{4r} = \small
\left(
\begin{array}{cc}
 0 & {\epsilon}_{2r}   \\
 {\epsilon}_{2r} & 0
\end{array}
\right)
\end{equation}
and we denote by $ M_{2s}$ a matrix of $\sym(2s)$.
We get a matrix of $\so(4rs, 4rs)$ by setting
\begin{equation}\label{pipB}
\left(
\begin{array}{c|c}
 A_{2r} \otimes {\mathbf 1}_{2s} +  {\mathbf 1}_{2r} \otimes  M_{2s} &
 B_{2r} \otimes  {\mathbf 1}_{2s}  \\
\hline
 C_{2r} \otimes  {\mathbf 1}_{2s}  &  -   A^T_r \otimes {\mathbf 1}_{2s} -  {\mathbf 1}_{2r} \otimes
 M^T_{2s}
\end{array}
\right)
\end{equation}
where the invariant tensor is now
 $$\epsilon_{4rs} = \small
\left(
\begin{array}{cc}
 0 & {\epsilon}_{2r} \otimes \epsilon_{2s}   \\
 {\epsilon}_{2r} \otimes \epsilon_{2s}  & 0
\end{array}
\right)\,.$$
Now we merge everything into a representation of $\so(2pq + 4rs, 2 pq + 4 rs)$
as follows
\begin{scriptsize}
\begin{equation}\label{pipC}
\left(
\begin{array}{c|c|c|c}
A_p \otimes {\mathbf 1}_{2q} +  {\mathbf 1}_p \otimes M_{2q}  &
0 &
B_p \otimes  {\mathbf 1}_{2q}  & 0\\
\hline
0 &
 A_{2r} \otimes {\mathbf 1}_{2s} +  {\mathbf 1}_{2r} \otimes  M_{2s} &
 0 &
 B_{2r} \otimes  {\mathbf 1}_{2s}  \\
\hline
C_p \otimes  {\mathbf 1}^T_{2q}  &
0 &
  -  A^T_p \otimes {\mathbf 1}_{2q} -  {\mathbf 1}_p \otimes  M^T_{2q}  &
0\\
\hline
0 &
 C_{2r} \otimes  {\mathbf 1}^T_{2s} &
0 &
-   A^T_{2r} \otimes {\mathbf 1}_{2s} -  {\mathbf 1}_{2r} \otimes
 M^T_{2s}
\end{array}
\right)\nonumber
\end{equation}
\end{scriptsize}
In this way we can construct the GZ kinetic term given the matrix (\ref{pipC}) for the
bosonic fields whose duality is now described by the
$$\frac{\SO( 2 pq + 4 rs, 2 pq + 4 rs)}{\SO(2pq + 4rs) \times \SO(2 pq + 4 rs)}\,.$$
This is the kinetic term for the bosonic fields of the sigma model.
The next step is to construct the duality for the fermion kinetic terms, namely we
have to embed the bosonic subgroups as follows
\begin{equation}\label{orpC}
\SO(p,p) \times {\rm Sp}(4r) \times \SO(q,q) \times {\rm Sp}(2s) \longmapsto
{\rm Sp}( 4ps + 8 qr)\,.
\end{equation}

However, it is rather straightforward to obtain the complete
solution for the embedding of the supergroups
\begin{equation}\label{supA}
\OSp(p,p|4r) \times \OSp(q,q|2s) \longmapsto \OSp( 2 pq + 4 rs, 2 pq + 4 rs | 4 p s + 8 q r )
\end{equation}
The symplectic algebra in the first factor $\sym(4r)$ must satisfy a condition on the rank, which has to be multiple of two due to the present decomposition.

A supermatrix $\widehat{\mathcal M}$ of $\osp(p,p|4r)$ satisfies
the condition $\widehat{\mathcal M}^{ST} \Omega + \Omega \widehat{\mathcal M}=0$, where
\begin{equation}\label{supAA}
\Omega = \left(
\begin{array}{cc}
 \eta_{2p} & 0    \\
 0 & \epsilon_{4r}
\end{array}
\right)\,, \quad \quad  \epsilon_{4r} = \left(
\begin{array}{cc}
 0 & \epsilon_{2r}    \\
 \epsilon_{2r} & 0
\end{array}
\right)\,,
\end{equation}
Decomposed in blocks $\widehat{\mathcal M}_{2p,4r}$ appears as follows
\begin{equation}\label{supAB}
\widehat{\mathcal M}_{2p,4r}= \left(
\begin{array}{c|c||c|c}
 A_p & B_p  &  - \delta^T \epsilon_{2r} & - \beta^T \epsilon_{2r} \\
 \hline
 C_p & - A^T_p  & - \gamma^T \epsilon_{2r}  & - \alpha^T \epsilon_{2r}\\
 \hline\hline
\alpha  & \beta  & {A}_{2r}  & { B}_{2r} \\
\hline
\gamma  & \delta  & {C}_{2r}  & \epsilon_{2r} {A}^T_{2r} \epsilon_{2r}
\end{array}
\right)
\end{equation}
where $A_p, {A}_{2r}$ are arbitrary even matrices and
$\alpha, \delta$ are arbitrary odd matrices. $B_p$ and $C_p$ are
antisymmetric matrices, while ${B}_{2r}$ and ${C}_{2r}$ satisfy
$${B}_{2r}^T \epsilon_{2r} + \epsilon_{2r} {B}_{2r}=0\,, \quad\quad
{C}_{2r}^T \epsilon_{2r} + \epsilon_{2r} {C}_{2r}=0\,.$$
Finally, $\beta$ and $\gamma$ are arbitrary odd matrices.
Now, we perform a change of basis such that a generic supermatrix appear in the
left-upper block of the $\osp(p,p|4r)$ matrix as follows
\begin{equation}\label{supAC}
{\cal O}^{-1} \widehat{\mathcal M}_{2p,4r} {\cal O} =  \left(
\begin{array}{c|c||c|c}
 A_p & - \delta^T \epsilon_{2r}  & B_p  & - \beta^T \epsilon_{2r} \\
 \hline
\alpha&  { A}_{2r}    &  \beta   & { B}_{2r}  \\
 \hline\hline
C_p   & - \gamma^T \epsilon_{2r}  & - A^T_p & - \alpha^T \epsilon_{2r} \\
\hline
\gamma  & { C}_{2r} &  \delta  & \epsilon_{2r} { A}^T_{2r}  \epsilon_{2r}
\end{array}
\right)
\end{equation}
where the supermatrices
\begin{eqnarray}\label{supACB}
&&
\widehat{ A}_{p,2r} =\left(
\begin{array}{c|c}
 A_p & - \delta^T \epsilon_{2r} \\
\hline
\alpha&  { A}_{2r}
\end{array}
\right)\,,
\quad\quad
\widehat{B}_{p,2r} =\left(
\begin{array}{c|c}
 B_p & - \beta^T \epsilon_{2r} \\
\hline
\beta&  {B}_{2r}
\end{array}
\right)
\,, \\
&&
\widehat{C}_{p,2r} =
\left(
\begin{array}{c|c}
 C_p & - \gamma^T \epsilon_{2r} \\
\hline
\gamma&  { C}_{2r}
\end{array}
\right)\,, \quad\quad
\widehat{D}_{p,2r} = - \Omega^{-1}_{p,2r} \widehat{A}^{ST} \Omega_{p,2r} =
\left(
\begin{array}{c|c}
- A^T_p & - \alpha^T \epsilon_{2r} \\
\hline
\delta  &  \epsilon_{2r} {A}^T_{2r}  \epsilon_{2r}
\end{array}
\right)\nonumber
\end{eqnarray}
with
$$
\Omega_{p,2r} =
\left(
\begin{array}{c|c}
 {\bf 1}_p & 0   \\
 \hline
0  & \epsilon_{2r}
\end{array}
\right)\,, \quad\quad
{\cal O} =
\left(
\begin{array}{c|c|c|c}
 {\bf 1}_{p} & 0  & 0 & 0 \\
 \hline
 0 &  0 & {\bf 1}_p & 0 \\
 \hline
 0 & {\bf 1}_{2r}  & 0  & 0\\
 \hline
 0 & 0  & 0  & {\bf 1}_{2r}
\end{array}
\right)
$$

$\widehat{ A}_{p,2r}$ a supermatrix of ${\mathfrak gl}(p|2r)$
and two matrices $\widehat{ B}_{p,2r}$ and $\widehat{ C}_{p,2r}$
of $\osp(p|2r)$, respectively.  (Notice that
counting the parameters of the matrix $\widehat{M}$ we have: $p (2 p-1) + 2 r (1 + 4 r)$ bosons and
$ 8 p r$ fermions. They can be decomposed into $p^2 + 4r^2$ bosons and $2 pr$ fermions from
${\mathfrak gl}(p|2r)$ and $p(p-1) +  2r (2r +1)$ from the two matrices $\osp(p|2r)$).

To construct the embedding we multiply tensorially the matrix ${\cal O}^{-1} \widehat{M}_{2p,4r} {\cal O}$
with the matrix $\widehat{N}_{2q,2s}$ belonging to $\osp(q,q|2s)$ as follows
\begin{equation}\label{supAD}
\widehat{\Lambda}_{2m,4n}=\left(
\begin{array}{c|c}
  \widehat{ A}_{p,2r} \otimes {\bf 1}_{2q,2s} + {\bf 1}_{p,2r} \otimes \widehat{ N}_{2q,2s}&
  \widehat{ B}_{p,2r} \otimes {\bf 1}_{2q,2s}
   \\
   \hline
\widehat{ C}_{p,2r} \otimes  {\bf 1}_{2q,2s}
   &
 \widehat{ D}_{p,2r} \otimes {\bf 1}_{2q,2s} + {\bf 1}_{p,2r} \otimes \widehat{N}_{2q,2s}
\end{array}
\right)
\end{equation}

where $m = 2 pq + 4rs$ and $n = ps + 2rq$.
The matrix $\widehat\Lambda_{2m,4n}$ is an element of  $\osp(m,m|4n)$ in the basis where the metric
appears in the form
$$\Omega =
\left(
\begin{array}{c|c}
 0 & \Omega_{p,2r} \otimes \Omega_{2q,2s}   \\
\hline
\Omega_{p,2r} \otimes \Omega_{2q,2s}   &  0
\end{array}
\right)
$$

We have derived the complete embedding of the direct product of supegroups into a
matrix $\widehat{ \Lambda}_{2m,4n}$ of $\OSp(m,m|2n)$, and we are now in position to
compute the GZ formula (\ref{masterformula2}).

\subsection{An example}\label{sec6}

Let us consider a D=2 model in the Elhers frame described by the following supercoset
\begin{equation}\label{exaA}
\frac{{\rm OSp}(3,3|6)}{{\rm SO}(3)\times {\rm SO}(3) \times {\rm SO}(2) \times {\rm SU}(3)}\,.
\end{equation}
The bosonic subgroup of ${\rm OSp}(3,3|4)$ is $SO(3,3)$ and ${\rm Sp}(8)
$which are non-compact. The maximal compact subgroup is $\SO(3)^2 \times {\rm U}(3)$.
It also contains 36 fermions organized in a representation $({\bf 6}, {\bf 6})$.

The corresponding Matzner-Missner model is described by the following supercoset
We start from the coset
\begin{equation}\label{exaB}
\frac{{\rm \OSp(2,2|2)}}{{\rm SO}(2)\times {\rm SO}(2) \times  {\rm
SO}(2)} \times \frac{{\rm \OSp(1,1|4)}}{U(2)} \oplus
\frac{1}{2}\Big[{(\bf (4|2}), (0|0)) \otimes  ((0|0), {\bf
(2|4}))\Big]\,,
\end{equation}
where the first two factors describe the proper scalar and fermionic fields while the linear representation
define the bosonic and fermionic 0-forms. This example fits our discussion in the previous section for $p=1, q=2, r=s=1$.

The bosonic part of the sigma model supercoset is
\begin{equation}\label{exaC}
\frac{{\rm SO}(2,2)}{{\rm SO}(2) \times {\rm SO}(2)} \times {\rm SO}(1,1) \oplus  \frac{1}{2}({\bf 4},{\bf 2})\,,
\end{equation}
It has $(6 - 1 -1) + 1 + 4 = 9$ degrees of freedom which are the fields of
$SO(3,3) / SO(3) \times SO(3)$.
The other bosonic part is described by
 \begin{equation}\label{exaD}
\frac{{\rm Sp}(2)}{{\rm SO}(2)} \times \frac{{\rm Sp}(4)}{{\rm U}(2)}\oplus \frac{1}{2}({\bf 2}, {\bf 4})\,,
\end{equation}
 The counting of degrees of freedom gives $(3-1) + (10-4) + 4 = 12$ which coincides with the
 degrees of freedom
 of ${\rm Sp}(6)/SO(2)\times SU(3)$.
Now, we are ready to fuse them into a supergroups as in (\ref{exD})
where the factor $1/2$ divides the number of bosonic degrees of
freedom. Summing up the bosonic degrees of freedom we have $(6 -1
-1) + (3-1) + 1 + (10 -4) + 1/2 (4 \times 2 + 2\times 4)= 21$ which
are the bosonic degrees of freedom of the original coset
(\ref{exA}).  It is easy to check that also the fermionic degrees of
freedom work. There are 16 fermions in the coset $(4 \times 2 +
2\times 4)$ and 20 fermions in the linear representation.

The duality group is easily found
\begin{equation}\label{exaE}
\frac{{\rm OSp}(8,8| 20)}{{ \OSp}(8|10) \times {\OSp}(8|10)}\,,
\end{equation}
which has $(164|160)$ degrees of freedom. This contains the embedding of the isometry transformations and new dualities
that we are going to explore in the next section.


\section{Superdualities}\label{sec5}

Having exploited the structure of the sigma model with the embedding of the
supergroup $\OSp(m,m|4n)$, we clarify now the structure of the duality transformations.
In particular we are interested in the superdualities, namely those transformations
of the supergroup which are not in the embedded
${\cal G}$. These transformations generate new sigma models
related by dualities.

 Let us first recall some basic facts about dualities in the usual framework.
 The subgroup $O(m,m)$ is generated by the following
 three set of transformations
 \begin{enumerate}
\item The field redefinitions are generated by the matrix $A$ of $GL(m,\mathbb{R})$
and embedded into the $O(m,m)$ as follows
\begin{equation}\label{ooA}
g_A = \left(
\begin{array}{cc}
 A & 0     \\
0  & (A^T)^{-1}
  \end{array}
\right)
\end{equation}
\item The second set are the shifts of the $B$ field and they are embedded into
the duality group as follows
\begin{equation}\label{ooB}
g_B= \left(
\begin{array}{cc}
 {\bf 1} & \Theta     \\
0  & {\bf 1}
  \end{array}
\right)
\end{equation}
where $\Theta$ is an $m\times m$ antisymmetric matrix and ${\bf  I}$ is the identity matrix
in $m$ dimension. This subgroup is a parabolic subgroup.
\item The factorized dualities are given by
\begin{equation}\label{ooC}
g_D = \left(
\begin{array}{cc}
{\bf  1} - e^i  & e^i     \\
e^i  & {\bf 1} - e^i
  \end{array}
\right)
\end{equation}
where $e^i$ is a diagonal matrix will all entries equal to zero except $\delta_{ii}$
(therefore it satisfies $e^i e^i = e^i$).
\end{enumerate}
It can be easily checked that these transformations generate the complete group and
in particular they are enough to describe all duality transformations. Combining the inversions
with the parabolic subgroup of point $(b)$, we generate a second $SO(m)$. Therefore, the total
number of parameters is $m^2 + m(m-1) = m (2m -1)$ which are the parameters of the group
$SO(m,m)$. The coset $SO(m,m)/SO(m)\times SO(m)$ decribes $m^2$ moduli of the kinetic terms
for the bosons in the action (\ref{d2generlag-super}).

Let us now move to the fermionic part of the action. We have already established that
the duality group is $Sp(4n)$ and therefore, we would like to repeat also in the present context
the analysis
\begin{enumerate}
\item Field redefinitions are generate by a matrix $A$ of $GL(2n,\mathbb{R})$
and they are embedded into the $Sp(4n, \mathbb{R})$ as follows
\begin{equation}\label{ssA}
g_A = \left(
\begin{array}{cc}
 A & 0     \\
0  & (A^T)^{-1}
  \end{array}
\right)
\end{equation}
\item The second set are the shifts of the WZ terms by a symmetric matrix and
they are embedded into the duality group as follows
\begin{equation}\label{ssB}
g_B= \left(
\begin{array}{cc}
 {\bf 1} & \Theta     \\
0  & {\bf 1}
  \end{array}
\right)
\end{equation}
where $\Theta$ is a symmetric matrix and ${\bf  1}$ is the identity matrix
in $2n$ dimension.
\item The factorized dualities are given by
\begin{equation}\label{ssC}
g_D = \left(
\begin{array}{cc}
{\bf  1} - e^i  & - e^i     \\
e^i  & {\bf 1} - e^i
  \end{array}
\right)
\end{equation}
where $e^i$ is a diagonal matrix will all entries equal to zero except $\delta_{ii}$
(therefore it satisfies $e^i e^i = e^i$). Notice the minus sign in the right-upper corner.
This is needed in order to be embedded in the duality group.
\end{enumerate}
It can be checked that these transformations generate the entire duality group.
Indeed the invertions combined with the shfits at point $(b)$ generate a second symplectic
subgroup ${\rm Sp}(2n)$. In total one has $(2 n)^2 + 2 n(2 n +1)$ parameters which generates the
total number of parameters of ${\rm Sp}(4n)$. Notice that the coset ${\rm Sp}(4n)/{\rm Sp}(2n)\times
{\rm Sp}(2n)$ has dimension $(2 n)^2$ and it describes all possible kinetic terms for the fermions.
(A way to interpret those terms in the context of string theory is to compare them with the pure spinor
formulation of string theory. From that point of view, the kinetic terms of fermions can be identified
with the RR fields).

Notice however, that the transformations generated by $\SO(m,m)$ and ${\rm Sp}(4n)$
are bosonic transformations. Therefore, we have to explore the transformations generated by the fermionic dualities.

Then, finally we have to cast everything in the supergroup. Therefore, we have to
analyze the off-diagonal supermatrices to see if they are still dualities of the action.
In that case the complete moduli space is captured by the supercoset
\begin{equation}\label{ooD}
\frac{OSp(m,m| 4n) }{OSp(m|2n) \times OSp(m|2n)}
\end{equation}
which has $m^2 + (2n)^2$ bosonic components and $4 n m$ fermionic components.

The supergroup $\OSp(m,m|4n)$ can be decomposed into $GL(m|2n)$ (which has
$m^2 + (2n)^2$ bosonic parameters and $4 nm$ fermions) plus two copies of $\OSp(m|2n)$
with $m(m-1)/2 + n(2n+1)$ bosonc and $2 nm$ fermions. The present discussion uses the
form of the supermatrices presented above where we have decomposed the matrix
$(2m + 4n)\times (2m+4n)$ into $(m + 2n)\times (m+ 2n)$ supermatrix blocks.

So, according to the previous
dsicussion, we have:
\begin{enumerate}
\item The field redefinitions are
generated by a generic supermatrix $\widehat{\cal A}$ of $GL(m|n)$ and
are embedded into the supergroup  are follows
\begin{equation}
\label{gnepA}
g_A =
\left(
\begin{array}{c|c}
 \widehat{\mathcal A} & 0    \\
\hline
0   &  {\bf \Omega}_{(m+2n)\times (m+2n)} (\widehat{\mathcal A}^T)^{-1}  {\bf \Omega}_{(n+2m)\times (n+2m)}^T
\end{array}
\right)
\end{equation}
where the matrix $ {\bf \Omega}_{(n+2m)\times (n+2m)}$ is defined in (\ref{embeddusmatra2}).
\item Shift by an orthosymplectic matrix $\widehat{\mathcal B}$ described by the
parabolic subgroup
\begin{equation}
\label{gnepAB}
g_B =
\left(
\begin{array}{c|c}
 {\bf 1}_{m+2n} & \widehat{\mathcal B}_{(m+2n)\times (m+2n)}   \\
\hline
0   &
{\bf 1}_{m+2n}
\end{array}
\right)
\end{equation}
The matrix $\widehat{\mathcal B}_{(m+2n)\times (m+2n)}$ is an element of the
orthosymplectic subgroup $\OSp(m|2n)$.
\item
The inversions are purely fermionic transformations and are obtained by the combination
of the orthogonal ones (\ref{ooC}) and the symplectic ones (\ref{ssC}).
\end{enumerate}
By combining the parabolic subgroup at point $(b)$ with the inversions, one finds
the other parabolic subgroup $\OSp(m|2n)$ and therefore the above transformations
generate the complete duality group as in the bosonic cases.


\subsection{Fermionic Inversions}

As the last section, we present a specific example obtained from duality which has an interesting interpretation
from string theory point of view.

We start fromhe simplest example of sigma model with the choice for couplings (\ref{grasimma})
\begin{equation}\label{exA}
\Gamma_{\Sigma\Lambda}(\Phi) = \Big( \delta_{\a\b}, 0, 0 \Big)\,,
\quad\quad
\Theta_{\Sigma\Lambda}(\Phi) = \Big( 0, 0, \delta_{\bar\a\bar\b}, \Big)\,.
\end{equation}
This choice corresponds to a sigma model with a simple kinetic term for the boson fields $\Pi^\alpha$
and a WZ term for the fermions $\Pi^{\bar\alpha}$. This situation resembles the starting quadratic action
for Green-Schwarz string theory (or the Pure Spinor Formulation) on a flat background. The bosonic
fields are identified with the coordinates $x^m$ and the fermions are identified with the superspace coordinates.
Notice that in order that the WZ term produces a quadratic term in the fermions one needs a non-trivial
background (see for example \cite{Berkovits:1999zq}). The matrix $\widehat {\cal M}$ has the form
\begin{equation}\label{exB}
\widehat {\cal M} =
\left(
\begin{array}{ccc}
  \delta_{\alpha \beta} &  0 \\
 0 &       \delta_{\bar\alpha \bar\beta}
\end{array}
\right)
\end{equation}
We perform a duality transformation with only fermionic parameters. Since we need the group element
which satisfies the equation
\begin{equation}\label{exC}
\left(
\begin{array}{c|c}
0  & \widehat{\cal B}  \\
 \hline
 \widehat{\cal C}  & 0
\end{array}
\right)^{ST}
\left(
\begin{array}{c|c}
 0 & \Omega \\
 \hline
\Omega  & 0
\end{array}
\right)
\left(
\begin{array}{c|c}
0  & \widehat{\cal B}  \\
 \hline
 \widehat{\cal C}  & 0
\end{array}
\right) =
\left(
\begin{array}{c|c}
 0 & \Omega \\
 \hline
\Omega  & 0
\end{array}
\right)
\end{equation}
we get that
$ \widehat{\cal C}^{ST} = \Omega \,  \widehat{\cal B}^{-1} \Omega^{-1}$. Therefore, we use the duality transformation
given by
\begin{equation}\label{exD}
\Lambda =
\left(
\begin{array}{cc}
0  &  \widehat{\cal B} \\
 \Omega \big( \widehat{\cal B}^{ST}\big)^{-1} \Omega^{-1}& 0
\end{array}
\right)
\end{equation}
The matrix  $\widehat{\cal B}$ is a supermatrix which must be inverted in order to get the contribution in the left-bottom quadrant.
To simplify the example, we consider a supermatrix $(1|2) \times (1|2)$ with the entries
\begin{equation}\label{exE}
\widehat{\cal B}=
\left(
\begin{array}{cc}
 b & \beta \\
\bar\beta  & \bar b
\end{array}
\right)\,,
\quad\quad
\widehat{\cal B}^{-1}=
\left(
\begin{array}{cc}
 b^{-1} + b^{-1} \beta \mathbf{B} \bar \beta& - \beta \mathbf{B} \\
- \mathbf{B}    \bar \beta & b \, \mathbf{B}
\end{array}
\right)\,,
\end{equation}
where $b \in \mathbb{R}$, $\beta, \bar \beta$ are vectors (with fermionic components) and
$\bar b$ and $\mathbf{B} = ({b \bar b - \bar \beta \beta})^{-1}$ are $2\times 2$ matrices.
Computing the supertranspose, we get
\begin{equation}\label{exF}
\Big(\widehat{\cal B}^{ST}\big)^{-1}=
\left(
\begin{array}{cc}
 b^{-1} + b^{-1} \beta \mathbf{B} \bar \beta& - \bar\beta^T \mathbf{B}^T \\
\mathbf{B}^T    \beta^T & b \, \mathbf{B}^T
\end{array}
\right)\,,
\end{equation}
Since we are interested into the inversion transformations due to the fermionic parameters
we compute the expression by performing an $\OSp$-transformation on $\widehat{\cal M}$
\begin{equation}\label{exG}
\widehat{\cal M} \rightarrow  \Omega^{-1} ( \widehat{\cal C} + \widehat{\cal D} \widehat{\cal M})
( \widehat{\cal A} + \widehat{\cal B} \widehat{\cal M})^{-1} =
\Omega^{-1} \widehat{\cal C}  \widehat{\cal B}^{-1} \widehat{\cal M}^{-1}
\end{equation}
and inserting the result in (\ref{exD}), we get
\begin{equation}\label{exH}
\widehat{\cal M} \rightarrow   \big(\widehat{\cal B}^{ST}\big)^{-1}\,  \Omega^{-1} \, \widehat{\cal B}^{-1} \widehat{\cal M}^{-1}
\end{equation}
In addition, if we set $\bar\beta =0, b=1, \bar b= {\bf 1}_2$ to simplify (without loosing in generality)
the result, this yields
\begin{equation}\label{exI}
\widehat{\cal M} =
\left(
\begin{array}{cc}
1  & - \beta \\
 \beta^T  & {\bf 1}_2 + \beta^T \beta
\end{array}
\right)\, \widehat{\cal M}^{-1} = \left(
\begin{array}{cc}
1  & -\beta \\
\beta^T  & {\bf 1}_2 + \beta^T \beta
\end{array}
\right)\,.
\end{equation}
The duality transformation has produced two additional couplings: the coupling between bosons $\Pi^\a$ and fermions $\Pi^{\bar \a}$ and
the coupling fermion-fermion. The first one has a fermionic nature since it must provide a fermionic coupling. Notice that one can be
tempted to interpret this coupling as induced by the target-space gravitinos represented here by the vector $\beta$. On the other hand,
the coupling fermion-fermion produces a new term in the kinetic term --
notice that the matrix $\beta^T \beta$ is antisymmetric due to the fermionic nature
or $\beta$ -- which can be interpreted as RR coupling.\footnote{In order to identify correctly the
RR fields in the fermion-fermion couplings one has to use the Pure Spinor Formulation of String Theory. Following
the discussion given in \cite{bebbe}, one way to see these coupling is indeed the contribution at the quadratic level of a
RR field in the supergravity background.}

The duality transformation of the form (\ref{exG}) are discussed for the first time in
works \cite{Berkovits:2008ic,Beisert:2008iq}. Here, we see that they enter our scheme completely
and indeed our derivation could produce
other non-trivial background to be studied.

\section*{Conclusions and outlook}

We have formulated in full generality the theory of superdualities. We have found that the correct construction
leads to the orthosymplectic dualitty supergroup $\OSp(m,m|4n)$. We have also insisted on the interpretation
of the sigma model as a model for superstrings on background with RR fields and with fermions. This can work in D=2, but
the interpretation in higher dimensions is puzzling and we have not found a proper interpretation from supergravity compactifications
except that in the case of twisted supersymmetry where some gravitinos are interpreted as ghosts. In addition, it has not be
explored the implications of the superdualities at the level of physical meaningful models. Finally,
the application of or formalism to double geometry and T-folds \cite{Hull:2004in} and double field theory \cite{Hull:2009mi} might be
quite interesting.


  \end{document}